\pdfoutput=1
\documentclass[fleqn,usenatbib,useAMS]{mnras}

\usepackage{graphicx}	
\usepackage{amsmath}	
\usepackage{amssymb}	
\usepackage{multicol}        
\usepackage{bm}		
\usepackage{pdflscape}	
\usepackage{epstopdf}
\usepackage[T1]{fontenc}
\usepackage{ae,aecompl}
\usepackage{physics}
\usepackage[dvipsnames]{xcolor}

\newcommand{\rev}[1]{\textcolor{black}{#1}}

\newcommand{\beq}{\begin{equation}}
\newcommand{\beqa}{\begin{eqnarray}}
\newcommand{\eeq}{\end{equation}}
\newcommand{\eeqa}{\end{eqnarray}}

\newcommand{\simgt}{\lower.5ex\hbox{$\; \buildrel > \over \sim \;$}}
\newcommand{\simlt}{\lower.5ex\hbox{$\; \buildrel < \over \sim \;$}}

\newcommand{\bv}{{\bf v}}
\newcommand{\bx}{{\bf x}}
\newcommand{\ba}{{\bf a}}

\usepackage[T1]{fontenc}
\usepackage{ae,aecompl}

\usepackage{newtxtext,newtxmath}

\newcommand{\Von}{von}
\newcommand{\Van}{van}

\title[Modelling SIDM substructures~I]{
Modelling self-interacting dark matter substructures~I:~
Calibration with N-body simulations of a Milky-Way-sized halo and its satellite
}

\author[M. Shirasaki et al.]
{
Masato Shirasaki$^{1,2}$\thanks{Contact e-mail: \href{mailto:masato.shirasaki@nao.ac.jp}{masato.shirasaki@nao.ac.jp}},
Takashi Okamoto$^{3}$,
and Shin'ichiro Ando$^{4,5}$
\\
$^{1}$National Astronomical Observatory of Japan, Mitaka, Tokyo 181-8588, Japan \\
$^{2}$The Institute of Statistical Mathematics, Tachikawa, Tokyo 190-8562, Japan \\
$^{3}$Faculty of Science, Hokkaido University, N10 W8, Kitaku, Sapporo, Hokkaido 060-0810 Japan\\
$^{4}$GRAPPA Institute, University of Amsterdam, 1098 XH Amsterdam, The Netherlands\\
$^{5}$Kavli Institute for the Physics and Mathematics of the Universe (WPI), University of Tokyo, Chiba 277-8583, Japan
}

\pubyear{2022}

\begin{document}
\label{firstpage}
\pagerange{\pageref{firstpage}--\pageref{lastpage}}
\maketitle

\begin{abstract}
We study evolution of single subhaloes with their masses of $\sim10^9 M_\odot$
in a Milky-Way-sized host halo for self-interacting dark matter (SIDM) models.
We perform dark-matter-only N-body simulations of 
dynamical evolution of individual subhaloes orbiting its host
by varying self-scattering cross sections (including a velocity-dependent scenario), 
subhalo orbits, and internal properties of the subhalo.
We calibrate a gravothermal fluid model to predict time evolution in spherical mass density profiles of isolated SIDM haloes with the simulations.
We find that tidal effects of SIDM subhaloes can be described with a framework developed for the case of collision-less cold dark matter (CDM), but a shorter typical time scale for the mass loss due to tidal stripping is required to explain our SIDM simulation results.
As long as the cross section is less than $\sim10\, \mathrm{cm}^2/\mathrm{g}$
and initial states of subhaloes are set within a $2\sigma$-level scatter at redshifts of $\sim2$ predicted by the standard $\Lambda$CDM cosmology, 
our simulations do not exhibit a prominent feature of gravothermal collapse 
in the subhalo central density for 10 Gyr. 
We develop a semi-analytic model of SIDM subhaloes in a time-evolving density core of the host with tidal stripping and self-scattering ram pressure effects.
Our semi-analytic approach provides a simple, efficient and physically-intuitive prediction of SIDM subhaloes, but further improvements are needed to account for baryonic effects in the host
and the gravothermal instability accelerated by tidal stripping effects.
\end{abstract}





\section{Introduction}

An array of astronomical observations has established 
a concordance cosmological model, referred to 
as $\Lambda$ Cold Dark Matter ($\Lambda$CDM) model. 
The $\Lambda$CDM model requires the presence of invisible mass components in the Universe
to explain the current observational data.
The nature of such ``dark'' matter is still uncertain.
Because dark matter plays an essential role in formation and evolution of cosmic large-scale structures, the observations of large-scale structures have constrained the cosmic abundance of dark matter in the Universe \citep[e.g.][]{2020A&A...641A...6P, 2021PhRvD.103h3533A}, free-streaming effects induced by thermal motion of dark matter particles \citep[e.g.][]{2016JCAP...08..012B, 2020JCAP...04..038P},
non-gravitational scattering of baryons and dark matter \citep[e.g.][]{2014PhRvD..89b3519D, 2018PhRvD..97j3530X}, 
electrically charged dark matter \citep[e.g.][]{2017PhRvD..95b3502K},
and annihilation and decay processes of dark matter particles \citep[e.g.][]{2015JCAP...05..024A, 2016PhRvD..94f3522S, 2017PhRvD..95b3010S, 2021JCAP...12..015K}.
So far, all constraints by the large-scale structures 
indicate that gravitational interactions are dominant in the growth of dark matter density, dark matter does not interact with ordinary matter and/or electromagnetic radiation,
and its thermal motion is negligible.

Although the $\Lambda$CDM model has provided an excellent fit to the observational data on length scales longer than $\sim10\, \mathrm{Mpc}$, 
it remains unclear if the model can be compatible with observations at smaller scales \citep[e.g.][for a review]{2017ARA&A..55..343B}.
Self-interacting dark matter (SIDM) has been proposed as a solution 
for the small-scale challenges to 
the $\Lambda$CDM model \citep[e.g.][]{2000PhRvL..84.3760S}. 
Elastic self-interactions among dark matter particles can lead to formation of a cored density profile, that is preferred by observations of galaxies and galaxy clusters.
After its proposal, numerical simulations have played a central role to improve 
our understanding of the structure formation in the presence of dark matter self-interactions, 
whereas particle physics models have been proposed to realise the SIDM preferred by some astronomical observations \citep[e.g.][for a review]{2018PhR...730....1T}.

Recently, \citet{2015MNRAS.452.3650O} 
found that rotation curves of observed spiral galaxies 
exhibit a diversity at their inner regions.
This diversity problem appears to conflict with the $\Lambda$CDM prediction, but 
it can be explained within a SIDM framework \citep[e.g.][]{2017PhRvL.119k1102K, 2019PhRvX...9c1020R, 2020JCAP...06..027K}.
Nevertheless, it would be worth noting that the SIDM solution to the diversity problem 
depends on the sampling of halo concentration as well as co-evolution of dark matter with baryons \citep[e.g.][]{2017MNRAS.468.2283C, 2020MNRAS.495...58S, 2021MNRAS.507..720S}. 

Satellite galaxies in the Milky Way (denoted as MW satellites) are 
promising targets for robustly constraining the SIDM scenarios.
The MW satellites are expected to be dominated by dark matter, and 
their dark matter contents would be less affected by possible baryonic effects 
inside the satellites.
\citet{2018NatAs...2..907V}
 examined the cross section of dark matter self-interactions with kinematic observations of MW dwarf spheroidals, 
but their modelling of SIDM density profiles does not include tidal effects from the host.
A similar investigation has been done for less massive satellites known as ultra-faint dwarf galaxies in \citet{2021PhRvD.103b3017H}.
\citet{2019MNRAS.490..231K} pointed out an anti-correlation between 
the central dark-matter densities of the bright MW satellites and 
their orbital peri-center distances inferred from Gaia data. 
The anti-correlation can be explained by a SIDM model \citep[e.g.][]{2021MNRAS.503..920C},
while a more careful modelling of the kinematic observations leads that the $\Lambda$CDM predictions can explain the anti-correlation \citep[e.g.][]{2020ApJ...904...45H}

High-resolution numerical simulations provide a powerful means of 
predicting the MW satellites in the presence of dark matter self-interactions 
\citep[e.g.][]{2019PhRvD.100f3007Z, 2022PhRvD.105b3016E, 2022arXiv220310104S} 
and the interplay with baryonic effects \cite[e.g.][]{2019MNRAS.490.2117R, 2020MNRAS.498..702L, 2021MNRAS.504.3509O}.  
However, numerical simulations can suffer from resolution effects
and are commonly expensive to scan a wider range of parameters of interest.
In practice, we need to account for various modelling uncertainties (e.g. possible baryonic effects and galaxy-halo connections) as well as several observational systematic effects to place a meaningful constraint of the nature of dark matter with 
the observations of the MW satellites \citep[e.g.][]{2021PhRvL.126i1101N, 2021arXiv210609050K}.
Looking towards future measurements in wide-field spectroscopic surveys 
\citep[e.g.][]{2014PASJ...66R...1T},
an efficient semi-analytic modelling of the MW satellites in the presence of dark matter self-interactions is highly demanded.

In this paper, we aim at developing 
a semi-analytic model of 
the SIDM satellite haloes (denoted as subhaloes) 
in a MW-sized host halo.
For this purpose, we perform a set of (dark-matter-only) N-body simulations of halo-subhalo mergers by varying the self-interacting cross sections, subhalo orbits, and internal properties of the subhaloes at their initial state.
For comparisons, we formulate a simple semi-analytic model of the SIDM subhaloes accreting onto the host halo based on previous findings for the collision-less dark matter \citep[e.g.][]{2019MNRAS.490.2091G, 2021MNRAS.502..621J}.
We then calibrate our semi-analytic model with the idealised N-body simulations 
and assess its limitation.
Our analysis would make an important first step toward 
a more precise modelling of the SIDM subhaloes, as well as improve our physical 
understanding of evolution of the SIDM subhaloes.

The rest of this paper is organised as follows. 
We describe our N-body simulations in Section~\ref{sec:simulation}.
Next, we summarise our semi-analytic model of the SIDM subhaloes in Section~\ref{sec:model}.
Section~\ref{sec:results} presents the key results, 
whereas we discuss the limitations of our analysis in Section~\ref{sec:limitations}.
Finally, concluding remarks are provided in Section~\ref{sec:conclusions}.
In the following, $\ln$ represents the natural logarithm.
Throughout this paper, we adopt $\Lambda$CDM cosmological parameters below;
the average cosmic mass density $\Omega_\mathrm{m} = 0.315$,
the cosmological constant $\Omega_\Lambda = 1-\Omega_\mathrm{m} = 0.685$,
the average baryon density $\Omega_\mathrm{b} = 0.0497$,
the present-day Hubble parameter $H_0 = 100h = 67.3\, \mathrm{km/s/Mpc}$,
the spectral index of the power spectrum of primordial curvature perturbations $n_s=0.96$,
and the linear mass variance within $8\, \mathrm{Mpc}/h$ being $\sigma_8=0.80$.
Those parameters are consistent with statistical analyses of cosmic microwave backgrounds in \citet{2020A&A...641A...6P}.
If necessary, we compute the critical density of the universe as 
$\rho_\mathrm{crit,z} = 2.775\times10^{11}\, [\Omega_\mathrm{m}(1+z)^3+\Omega_\Lambda]\, h^2\, M_\odot/\mathrm{Mpc}^{3}$,
where $z$ is a redshift.

\section{Simulations}\label{sec:simulation}

In this paper, we perform N-body simulations of idealised minor mergers
to study evolution of single subhaloes in an external potential by a host halo for SIDM models. This section summarises how to set initial conditions of our N-body simulations, our N-body simulation code, and physical parameter sets adopted in our simulations.

\subsection{Initial conditions}\label{subsec:inital_condition}

We assume that either host halo or subhalo at its initial state follows 
a spherical Navarro-Frenk-White \citep[NFW;][]{1997ApJ...490..493N} density profile.
At a given halo-centric radius $r$, the NFW profile is given by
\beqa
\rho_\mathrm{NFW}(r) = \frac{\rho_s}{(r/r_s)(1+r/r_s)^{2}}, \label{eq:NFW}
\eeqa
where $\rho_s$ and $r_s$ represent the scaled density and radius, respectively.
The scaled density and radius can be related to a spherical over-density mass as
\beqa
M_{\Delta} = \frac{4\pi}{3}\, \Delta\, \rho_{\mathrm{crit}, z}\, r_{\Delta}^3 = \int_0^{r_\Delta}\, 4\pi r^2 \mathrm{d}r\, \rho_\mathrm{NFW}(r),
\eeqa
where $M_{\Delta}$ is the spherical over-density mass
and $r_\Delta$ is the corresponding halo radius.
Throughout this paper, we adopt a conventional mass definition with $\Delta = 200$.
The halo concentration is defined as $c=r_{200}/r_s$ and
a set of $c$ and $M_{200}$ can fully determine the NFW profile.
In the following, we use subscripts 'h' and 'sub' to indicate properties of 
the host- and sub-haloes, respectively.

For an initial condition of our N-body simulation, 
we fix the host halo mass, the halo radius and the scaled radius to 
$M_{200,\mathrm{h}} = 10^{12}\, M_\odot$, 
$r_{200,\mathrm{h}}=211\, \mathrm{kpc}$, and 
$r_{s,\mathrm{h}}=21.1\, \mathrm{kpc}$, respectively.
Note that the scaled density and radius of the host halo are set with the critical density at $z=0$.
For our fiducial case, we adopt 
$M_{200,\mathrm{sub}} = 10^{9}\, M_\odot$ and $c_\mathrm{sub}=6$ in the initial subhalo density,
but we vary $M_{200,\mathrm{sub}}$ and $c_\mathrm{sub}$ as necessary.
The initial subhalo concentration is set to be consistent with a model prediction 
in \citet{2015ApJ...799..108D} at $z=2$.
It would be worth noting that the redshift of $z=2$ provides a typical formation epoch of the 
$\sim10^{12} \, M_\odot$ halo at $z=0$ in the excursion set approach \citep{1991ApJ...379..440B, 1993MNRAS.262..627L}.
To keep a consistency with our choice of $c_\mathrm{sub}=6$, we determine $\rho_{s,\mathrm{sub}}$
and $r_{s,\mathrm{sub}}$ with the critical density at $z=2$.
Using different redshifts to define the initial density profiles of the host and subhalo
is a bit ambiguous, but our simulations do not contain accreting mass around the host
and there are no unique ways to realise a realistic situation as in cosmological simulations.
Because the outskirt region of the host halo is less important for orbital evolution of the subhalo, 
our simulations would be still useful to develop a better physical understanding
of orbiting SIDM subhaloes.

To generate isolated NFW host halo and subhalo,
we use a public code of {\tt MAGI} \citep{2018MNRAS.475.2269M}, assuming
that the NFW (sub)halo has an isotropic velocity distribution.
The code employs a distribution-function-based method so that the phase-space distribution 
of member particles in halos can be determined by energy alone.
To realise the system of particles in dynamical equilibrium with a sharp cut-off at $r\simeq r_{200}$,
we multiply the target NFW density profile with a function of $\mathrm{erfc}([r-r_{200}]/[2r_\mathrm{cut}])/2$,
where we adopt $r_\mathrm{cut} = 0.05\, r_\mathrm{200}$.
The number of particles is set to $10^7$ for the host halo, corresponding to the particle mass being $m_\mathrm{part}=10^{5}\, M_\odot$.
The convergence tests of our N-body simulations are summarised in Appendix~\ref{apdx:convergence_tests}.
We confirmed that our choice of the particle mass can provide converged results of subhalo mass loss with a level of $1\%$, and subhalo density profiles at $r/r_{s,\mathrm{sub}}\simgt 0.2$ 
within $10\%$ over 10 Gyr.

To specify the subhalo orbit, we introduce two dimensionless quantities $x_c$ and $\eta$.
In this paper, we express the angular momentum $L$ and the total energy $E$ of the orbiting subhalo as
\beqa
L &=& \eta r_c V_c \\
E &=& \frac{V^2_c}{2} + \Phi_\mathrm{NFW,h}(r_c),
\eeqa
where $r_c = x_c r_{200, \mathrm{h}}$,
$V_c = (G M_{200,\mathrm{h}}/r_c)^{1/2}$ is 
a velocity at the circular orbit when we treat host- and sub-haloes as isolated point particles,
and $\Phi_\mathrm{NFW,h}$ presents the gravitational potential by the host NFW profile \citep{2001MNRAS.321..155L}. 
The orbital period $T_r$ is then defined by
\beqa
T_r = \int_{r_p}^{r_a}\, \frac{\mathrm{d}r}{(2[E-\Phi_\mathrm{NFW,h}(r)]-L^2/r^2)^{1/2}},
\eeqa
where two radii $r_p$ and $r_a$ are given as a solution of the equation below:
\beqa
\frac{L^2}{r^2} + 2[\Phi_\mathrm{NFW,h}(r)-E] = 0.
\eeqa
The parameter $x_c$ controls the orbital period, whereas
$\eta$ determines the eccentricity in the subhalo orbit. 
We choose $x_c=0.5$ and $\eta=0.6$ as our baseline parameters, while we examine different values 
to test our semi-analytic model described in Section~\ref{sec:model}.
The baseline parameters provide $r_p=41.9\, \mathrm{kpc}$, $r_a=243\, \mathrm{kpc}$ and $T_r = 3.0\, \mathrm{Gyr}$ for our host halo.
For a given set of $x_c$ and $\eta$, we compute the initial (Cartesian) vectors of the
subhalo position and velocity with respect to the host halo
as $\bx_\mathrm{sub} = (r_a, 0, 0)$ and $\bv_\mathrm{sub} = (0, L/r_a, 0)$, respectively.
Note that the subhalo orbit is confined to the $x-y$ plane in our simulations.

\subsection{N-body simulations}

For a given initial condition of halo mergers, we evolve the system by solving gravitational and self-interactions among N-body particles.
To do so, we use a (non-cosmological) self-gravity mode of a flexible, massively-parallel, multi-method multi-physics code {\tt GIZMO} \citep{2015MNRAS.450...53H} for the gravitational interaction. 
Throughout this paper, we assume isotropic and elastic 
self-interaction processes in our simulations.

Our SIDM implementation follows the method in \citet{2017MNRAS.465..569R}.
In short, the rate with which a dark matter particle\footnote{A "particle'' here means a numerical element and should be distinguished from an SIDM particle of mass $m$.}, $i$, is scattered by other dark matter particles within the distance $h$ is given as:
\beq
{\cal R}_i = \qty(\frac{4 \pi}{3}h^3)^{-1} m_\mathrm{p}\sum_j \frac{\sigma(v_{ij})}{m} v_{ij}, 
\label{eq:SIDM-rate}
\eeq
where $m_\mathrm{p}$ is the mass of a dark matter particle as a numerical element, $v_{ij} = |\bm{v}_i - \bm{v}_j|$ is the relative speed between particles $i$ and $j$, and the sum is over all particles within the distance $h$ from the particle $i$.   
As in \citet{2017MNRAS.465..569R}, we apply a fixed value of $h$ to all particles. 
\rev{
The implementation with a constant $h$ has two advantages over one with a variable $h$ in accord with the local density.
As we discuss later, the symmetry between a pair of particles is important for the accurate scattering rate estimation. 
We also do not need expensive iterative loops when using a constant $h$, 
whereas the loops can become expensive for the adaptive $h$ to make the (effective) number of neighbouring particles within $h$ constant.  
}
We set $h = 2.8 \epsilon$, where $\epsilon$ is the Plummer equivalent force softening length and the gravitational force becomes Newtonian at $2.8 \epsilon$. 

From Eq.~(\ref{eq:SIDM-rate}), the probability of the particle, $i$, is scattered by one of 
its neighbours, $j$, within a distance $h$ during a time-step $\Delta t_i$ is 
\beq
P_{ij} = \frac{1}{2}\qty(\frac{4 \pi}{3}h^3)^{-1} m_\mathrm{p}
\frac{\sigma(v_{ij})}{m} v_{ij}  \Delta t_i. 
\eeq
We introduce the factor $1/2$ since a scatter event always involves a pair of particles. 
\rev{The prefactor of $1/2$ is justified only when 
the identical intersection radius of $h$ is adopted to every neighbour particle.
For an adaptive $h$, 
we may need to introduce symmetrization as is usually done in 
the smoothed particle hydrodynamics \citep[e.g.][]{2010ARA&A..48..391S}. 
}


\rev{
For a scattering event between particles $i$ and $j$, we update their velocities as follows: 
\begin{align*}
    \bm{u}_i &= \bm{v}_\mathrm{cm} + (v_{ij}/2) \hat{\bm{e}} \\
    \bm{u}_j &= \bm{v}_\mathrm{cm} - (v_{ij}/2) \hat{\bm{e}}, 
\end{align*}
where $\bm{u}_i$ and $\bm{u}_j$ are the post-scatter velocities of the particle $i$ and $j$, respectively, $\bm{v}_\mathrm{cm} = (\bm{v}_i + \bm{v}_j)/2$ is the centre-of-mass velocity, and $\hat{\bm{e}}$ is the randomly oriented unit vector.
}
\rev{
We have tested our SIDM implementation by counting the number of collisions of N-body particles in a spherical halo 
and observing post-scattering kinematics in a uniform background as in \citet{2017MNRAS.465..569R}, and confirmed it agrees with the analytic expectation}.

\rev{
In principle, a particle can scatter more than once in a single time-step, even if we employ a very short time step.
Multiple scatters in a single time step may introduce undesired numerical errors because the momentum kick from one scattering event affects the velocities of particles for any further scattering events.
To minimise possible numerical artifacts, we update the particle velocities immediately after setting relevant particles to scattering processes.
}

\rev{Running simulations on multiple processors with domain decomposition can cause a further complication because a particle can undergo scattering events among different computational domains.
To avoid any confusions, we first perform the SIDM calculation on the local domain where we can easily apply the immediate velocity update. 
When a particle is exported to other computational domains, the SIDM calculations are performed in the export destinations in the same manner as in the local domain. 
If an exported particle undergoes scattering events in two or more destinations or an exported particle scatters in one of the destinations and the same particle is scattered by an imported particle in the local domain, 
these scattering processes violate the energy conservation.
To reduce such bad scatters, we restrict the time-step $\Delta t_i$ to be smaller than $0.02/{\cal R}_i$ as often done in the literature \citep[e.g.][]{2012MNRAS.423.3740V}. 
We have confirmed that the above procedure does not introduce detectable numerical errors on the conservation of total energy and momentum in an isolated system.
}

%

\rev{To test our SIDM implementation, we evolved a cluster-sized isolated halo following a Hernquist profile at its inital state with the same simulation setup
as in \citet{2017PhDT.......206R}. We then compared our simulation results with one in \citet{2017PhDT.......206R}.
We found that the halo core evolution in our simulation provide a good fit to the results in \citet{2017PhDT.......206R}, demonstrating that the scattering of N-body particles is correctly implemented. The test results are summarised in Appendix~\ref{apdx:test_robertson}.}

The box size on a side is set to $1100\, \mathrm{kpc}$ so that the boundary of our simulation box can not affect the simulation results. 
We also adopt the gravitational softening length, in terms of an equivalent-Plummer value, $\epsilon$, as proposed in \citet{2018MNRAS.475.4066V};
\beqa
\epsilon = 0.05 \, r_{s, \mathrm{sub}}\left(\frac{N_\mathrm{sub}}{10^{5}}\right)^{-1/3}, \label{eq:soften_length}
\eeqa
where $N_\mathrm{sub}$ represents the number of N-body member particles in initial subhaloes
and is set to $N_\mathrm{sub}=10^4$ for our baseline run.
All simulations output particle snapshots with a fixed time-step of $0.1\, \mathrm{Gyr}$ and stop at $t=10\, \mathrm{Gyr}$.
At each snapshot, we define gravitational-bound particles in the subhalo 
with the iterative method in \citet{2018MNRAS.475.4066V}.

\subsection{Parameters}

\begin{table*}
 \caption{Summary of parameters in our N-body simulations of halo mergers.
 For all simulations in this paper, we fix the host halo mass $M_{200,\mathrm{h}}=10^{12}\, M_\odot$, the scaled radius (in the initial NFW density) $r_{s, \mathrm{h}}=21.1\, \mathrm{kpc}$, and the concentration $c_{\mathrm{h}}=10$.
 In every simulation, we evolve the orbit of an infalling subhalo for 10 Gyr.
 Note that our simulations allow a time evolution of the host halo density in accord with the thermalisation due to the self-scattering process of dark matter particles.
 In each cell, $M_{200,\mathrm{sub}}$ is the initial subhalo mass, $r_{s, \mathrm{sub}}$ is the scaled radius in the initial subhalo density,
 $c_\mathrm{sub}$ is the subhalo concentration at its initial state,
 $\sigma/m$ is the self-scattering cross section per unit mass,
 and $(x_c, \eta)$ present dimensionless orbital parameters described in Subsection~\ref{subsec:inital_condition}.
 }
 \label{tab:sim_params}
 \begin{tabular}{lccccc}
  \hline
  Name & 
  $M_{200, \mathrm{sub}}\, (M_\odot)$ &
  $r_{s, \mathrm{sub}}\, (\mathrm{kpc})$ &
  $c_\mathrm{sub}$ &
  $\sigma/m\, (\mathrm{cm}^2/\mathrm{g})$ & 
  $(x_c, \eta)$ \\
  \hline
  \hline
  Fiducial ($v$-independent $\sigma/m$) & & & & & \\
  \hline
  \hline
  CDM & $10^9$ & $1.68$ & 6 & 0 & $(0.5, 0.6)$  \\ 
  SIDM1 & $10^9$ & $1.68$ & 6 & 1 & $(0.5, 0.6)$ \\
  SIDM3 & $10^9$ & $1.68$ & 6 & 3 & $(0.5, 0.6)$ \\
  SIDM10 & $10^9$ & $1.68$ & 6 & 10 & $(0.5, 0.6)$ \\
  \hline
  \hline
  $v$-dependent $\sigma/m$ & & & & & \\
  \hline
  \hline
  vSIDM & $10^9$ & $1.68$ & 6 & Eq.~(\ref{eq:vdep_sigma_over_m}) & $(0.5, 0.6)$ \\
  \hline
  \hline
  Different orbits & & & & & \\
  \hline
  \hline
  SIDM1-diff-orbit & $10^9$ & $1.68$ & 6 & 1 & $(0.6, 0.05)$, $(0.6, 0.35)$, $(0.6, 0.65)$, $(0.6, 0.95)$ \\
    & & & & & $(0.9, 0.05)$, $(0.9, 0.35)$, $(0.9, 0.65)$, $(0.9, 0.95)$ \\
    & & & & & $(1.2, 0.05)$, $(1.2, 0.35)$, $(1.2, 0.65)$, $(1.2, 0.95)$ \\
    & & & & & $(1.5, 0.05)$, $(1.5, 0.35)$, $(1.5, 0.65)$, $(1.5, 0.95)$ \\
  \hline
  \hline
  Varied subhalo properties & & & & & \\
  \hline
  \hline
  High $c_\mathrm{sub}$ & $10^9$ & $0.842$ & 12 & 1 & $(0.5, 0.6)$ \\
  Low $c_\mathrm{sub}$ & $10^9$ & $3.36$ & 3 & 1 & $(0.5, 0.6)$ \\
  Large $M_\mathrm{sub}$ & $10^{10}$ & $4.38$ & 5 & 1 & $(0.5, 0.6)$ \\
  \hline
 \end{tabular}
\end{table*}

Table~\ref{tab:sim_params} summarises a set of parameters adopted in our N-body simulations. Most simulations assume that the SIDM cross section per unit mass $\sigma/m$ is independent of relative velocities between dark matter particles, but we also explore the impact of a velocity-dependent $\sigma/m$ by adopting effective-range theories in \citet{2020JCAP...06..043C}. 
To be specific, we adopt a velocity-dependent scenario as in \citet{2020JCAP...06..043C};
\beqa
\frac{\sigma}{m} = \frac{\sigma_0}{m} \left\{\left[1-\frac{1}{8}\frac{r_e}{a}\left(\frac{v}{v_0}\right)^2\right]^{2}
+\frac{1}{4}\left(\frac{v}{v_0}\right)^2\right\}^{-1}, \label{eq:vdep_sigma_over_m}
\eeqa
where we set $\sigma_0/m = 6.3\, \mathrm{cm}^2/\mathrm{g}$,
$a = 37.4\, \mathrm{fm}$,
$r_e = -748.9\, \mathrm{fm}$,
and $v_0 = 100\, \mathrm{km}/\mathrm{s}$
and those parameters provide a reasonable fit to the observational constraints of $\langle \sigma v \rangle/m$ at the average relative velocity of $\langle v \rangle=10-100\, \mathrm{km}/\mathrm{s}$ in \citet{2016PhRvL.116d1302K}.
This velocity-dependent model predicts that an effective cross section $\langle \sigma v \rangle/m/\langle v \rangle$ is found to be $1-6\, \mathrm{cm^2}/\mathrm{g}$ at the mass scale of $\sim 10^{9}\, M_\odot$, while the cross section becomes smaller than $\sim 0.1\, \mathrm{cm^2}/\mathrm{g}$ for a MW-sized halo.

Apart from our fiducial orbital parameters ($x_c=0.5$ and $\eta=0.6$), we also examine 16 different orbits in a range of $0.6 \le x_c \le 1.5$ and 
$0.05 \le \eta \le 0.95$. Note that the range of $x_c$ and $\eta$ is consistent with the cosmological N-body simulation in \citet{2015MNRAS.448.1674J}.
For the initial density profile of an infalling subhalo, we vary the halo concentration by a factor of $2$ or $1/2$ but fix subhalo mass to $M_{200, \mathrm{sub}}=10^{9}\, M_\odot$. The change of $c_\mathrm{sub}$ by a factor of $2$ or $1/2$ roughly covers a $2.5\sigma$-level difference in the halo concentration at the mass of $10^{9}\, M_\odot$ in cosmological simulations \citep[e.g.][]{2013ApJ...767..146I}.
As another test, we consider a more massive infalling subhalo 
with $M_{200,\mathrm{sub}}=10^{10}\, M_\odot$ and $c_\mathrm{sub}=5$.
As in Subsection~\ref{subsec:inital_condition}, the density profile for the $10^{10}\, M_\odot$ subhalo is set with the critical density at $z=2$.


\section{Model}\label{sec:model}

This section describes our semi-analytic model of 
orbital and dynamical evolution of an infalling subhalo in the presence of 
self-interactions of dark matter particles.
The model consists of three ingredients; 
(i) a time-evolving SIDM density profile in isolation (Subsection~\ref{subsec:gravothermal}),
(ii) the equation of motion of the subhalo including dynamical friction and ram-pressure-induced deceleration (Subsection~\ref{subsec:eq_motion}), 
and (iii) mass loss of the subhalo across its orbit (Subsection~\ref{subsec:mass_loss}).
In the Subsections~\ref{subsec:gravothermal}-\ref{subsec:mass_loss}, we first assume a velocity-independent cross section $\sigma/m$ for simplicity.
We then describe how to include the velocity-dependence of $\sigma/m$ in our model in Subsection~\ref{subsec:treatment_vdependent_case}.

\subsection{Gravothermal fluid model}\label{subsec:gravothermal}

In our model, we follow a gravothermal fluid model \citep[e.g.][]{2002ApJ...568..475B}
to predict spherical density profiles of isolated haloes.
The gravothermal fluid model assumes that 
SIDM consists of a thermally conducting fluid in quasistatic equilibrium 
and the system of interest is isotropic and spherically-symmetric.
At a given time of $t$ and halo-centric radius of $r$, 
dark matter particles have a mass density profile $\rho(r, t)$.
Their one-dimensional (1D) velocity dispersion $\sigma_v(r, t)$ is set 
by the hydrostatic equilibrium of ideal gas at each moment;
\beqa
\frac{\partial p(r, t)}{\partial r} = -\frac{G M(r, t)\, \rho(r,t)}{r^2}, \label{eq:hydro_eq}
\eeqa
where $p=\rho\, \sigma^2_v$ is an effective pressure, 
$M(r,t)$ is the enclosed mass within the radius of $r$ at $t$, 
and we impose the mass conservation of
\beqa
\frac{\partial M(r,t)}{\partial r} = 4\pi \, r^2\, \rho(r,t). \label{eq:mass_conserve}
\eeqa
The thermal evolution of the fluid is governed by Fourier’s law of thermal conduction 
and the first law of thermodynamics,
\beqa
\frac{L(r,t)}{4\pi r^2} &=& -\kappa \frac{\partial T(r,t)}{\partial r}, \label{eq:energy_flux}\\
\frac{\partial L(r,t)}{\partial r} &=& -4\pi r^2 p(r,t)\, \left(\frac{\partial}{\partial t}\right)_M\, \ln \left(\frac{\sigma^3_v(r,t)}{\rho(r,t)}\right), \label{eq:1stlaw_thermo}
\eeqa
where $L(r,t)$ is the luminosity through a sphere at $r$, 
$T(r,t)$ is a temperature defined as $k_B T = m \sigma^2_v$ 
($m$ is the particle mass and $k_B$ is the Boltzmann constant),
$\kappa$ is the thermal conductivity, and
the time derivative in the right hand side of Eq.~(\ref{eq:1stlaw_thermo}) is Lagrangian.

As discussed in \citet{2002ApJ...568..475B}, we adopt a single expression of Eq.~(\ref{eq:energy_flux}) by considering both the cases where the the mean free path between collisions is significantly shorter or larger than the system size,
\beqa
\frac{L}{4\pi r^2} &=& -\frac{3}{2}b_{*}\rho\sigma_v\left[\left(\frac{1}{\lambda}\right)+\left(\frac{b_{*} \sigma_v t_r}{C_{*}\, H_g^2}\right)\right]^{-1} \frac{\partial \sigma^2_v}{\partial r}, \label{eq:energy_flux_B02}
\eeqa
where $H_g \equiv \sqrt{\sigma^2_v/(4 \pi G\rho)}$ 
is the gravitational scale height of the system, 
$\lambda=(\rho \sigma/m)^{-1}$ is the collisional scale for the mean free path, 
$t_r \equiv \lambda/(a\sigma_v)$ is the relaxation time with a coefficient of order of unity being $a$, 
and we adopt $a= \sqrt{16/\pi}$ for hard-sphere scattering of particles with a Maxwell-Boltzmann velocity distribution \citep{1965fstp.book.....R}. 

In Eq.~(\ref{eq:energy_flux_B02}), we introduce two model parameters of $b_{*}$ and $C_{*}$.
In the limit of $\lambda \ll H_g$, the thermal conductivity is given by 
$\kappa \simeq (3/2) (k_B/m) b_{*} \rho \lambda^2 /(a t_r)$ and $b_{*}$ can be regarded as an effective impact parameter among particle collisions.
In the limit of $\lambda \gg H_g$, one finds $\kappa \simeq (3/2)(k_B/m)C_{*}\rho H^2_g / t_r$, reproducing an empirical formula of gravothermal collapse of globular clusters \citep{1980MNRAS.191..483L}.
As our baseline model, we adopt $b_{*}=0.25$ and $C_{*}=0.75$ as proposed in \citet{2011MNRAS.415.1125K}. 
By assuming the NFW halo at $t=0$, 
we then numerically solve Eqs.~(\ref{eq:hydro_eq}), (\ref{eq:mass_conserve}), (\ref{eq:1stlaw_thermo}) and (\ref{eq:energy_flux_B02}) with the method described in Appendix A of \citet{2020PhRvD.101f3009N} \citep[also see][]{2015ApJ...804..131P}.

We note that \citet{2011MNRAS.415.1125K} found the parameters of $b_{*}=0.25$ and $C_{*}=0.75$ to explain their N-body simulations of isolated haloes following a self-similar solution of the gravothermal fluid model in \citet{2002ApJ...568..475B}.
Hence, we validate the gravothermal fluid model with 
$b_{*}=0.25$ and $C_{*}=0.75$ for NFW haloes at $t=0$ 
by using our N-body simulations of isolated haloes.
The comparisons with the gravothermal fluid model and our simulation results
are summarised in Appendix~\ref{apdx:gravthermal_NFW}.
We find that a correction of the gravothermal fluid model is needed to explain our simulation results for initial NFW haloes with their mass of $M_{200} = 10^{12}\, M_\odot$ and concentration of $c=10$ 
in the range of $0.3 \simlt \sigma/m \, (\mathrm{cm}^2/\mathrm{g}) \simlt 30$ 
at $t \le 10\, \mathrm{Gyr}$.
The final model of density profiles of isolated SIDM haloes is then given by
\beqa
\rho_\mathrm{SIDM}(r, t) = \rho_\mathrm{gt}(r, t)
\frac{x^{\beta}+(1/2)^{\beta}}{\left(x+\gamma/2\right)^{\beta}}, \label{eq:final_iso_density}
\eeqa
where $\rho_\mathrm{gt}(r, t)$ is the gravothermal-fluid prediction with $b_{*}=0.25$ and $C_{*}=0.75$ and $x=r/(0.1 r_s)$ ($r_s$ is the scaled radius of the initial NFW halo). 
The two parameters $\beta$ and $\gamma$ in Eq.~(\ref{eq:final_iso_density}) depend on time as well as $\sigma/m$;
\beqa
\beta  &=& 0.275 \, \left[\log_{10}(t/t_0)-0.492\right]^{2} + 1.38, \label{eq:gv_param3}\\
\gamma &=& 0.493 \, (t/t_0)^{0.203}, \label{eq:gv_param4}
\eeqa
where we introduce a characteristic time scale of 
\beqa
t_0 &\equiv& \left(\sqrt{\frac{16}{\pi}} \rho_s \frac{\sigma}{m} \sqrt{4\pi G \rho_s r^2_s} \right)^{-1} \nonumber \\
&=& 1.29\, \mathrm{Gyr}\, \left(\frac{\sigma/m}{1\, \mathrm{cm^2}/{g}}\right)^{-1} \left(\frac{\rho_s}{5\times 10^{6}\, M_\odot \mathrm{kpc}^{-3}}\right)^{-3/2}
\left(\frac{r_s}{20\, \mathrm{kpc}}\right)^{-1}, \label{eq:t_0_SIDM}
\eeqa
and 
note that $\sqrt{4\pi G \rho_s r^2_s}$ in the above equation 
provides a characteristic velocity for the initial NFW haloes.
Our model has been calibrated with N-body simulations of isolated SIDM haloes 
with the specific initial NFW profile ($M_{200}=10^{12}\, M_\odot$, $r_{200} = 211\, \mathrm{kpc}$, $r_s = 21.1\, \mathrm{kpc}$ and $\rho_s = 5.72\times 10^{6}\, M_\odot\, \mathrm{kpc}^{-3}$), but 
we use Eq.~(\ref{eq:final_iso_density}) for any initial NFW profiles in the following.

\subsection{Orbital evolution}\label{subsec:eq_motion}

Assuming that the subhalo is not significantly deformed by tidal forces and self-interactions, 
we treat it as a point particle.
Under this point-mass approximation, we evaluate the orbit of the subhalo by solving 
the equation of motion \citep[e.g.][for the same approach]{2021arXiv210803243J, 2021MNRAS.502..621J}, 
\beqa
\frac{\mathrm{d}^2 \bx_\mathrm{sub}}{\mathrm{d}t^2}
= -\nabla \Phi_\mathrm{h} + \ba_\mathrm{DF} + \ba_\mathrm{RPd}, \label{eq:motion}
\eeqa
where $\Phi_\mathrm{h}$ is the gravitational potential of a SIDM host halo 
with its density following Eq.~(\ref{eq:final_iso_density}), 
$\ba_\mathrm{DF}$ represents the acceleration due to dynamical friction,
and $\ba_\mathrm{RPd}$ is the deceleration causing by the scattering process among 
escaping dark matter particles from the infalling subhalo and particles in the host halo \citep{2018MNRAS.474..388K}.

On the term of dynamical friction, we adopt the Chadrasekhar formula \citep{1943ApJ....97..255C} as
\beqa
\ba_\mathrm{DF} = -4 \pi G^2 \, M_\mathrm{sub}\, \rho_\mathrm{h}\, \ln \Lambda \, F_v(|\bv_\mathrm{sub}|)\, \frac{\bv_\mathrm{sub}}{|\bv_\mathrm{sub}|^3},
\eeqa
where we adopt an expression of the Coulomb logarithm as $\ln \Lambda = \xi\, \ln (M_\mathrm{h}/M_\mathrm{sub})$ with a fadge factor of $\xi$ being $\mathrm{min}(|\mathrm{d}\ln \rho_\mathrm{h}/\mathrm{d}\ln r|, 1) $ at $r=|\bx_\mathrm{sub}|$ as proposed in \citet{2006MNRAS.373.1451R}, 
and 
\beqa
F_v(v) = \mathrm{Erf}(y) - 2y \exp(-y^2)/\sqrt{\pi}
\eeqa
with $y=v/(\sqrt{2}\sigma_{v,\mathrm{h}})$ for an isotropic and Maxwellian host halo.
The velocity dispersion of $\sigma_{v,\mathrm{h}}$ is given by the solution of Eq.~(\ref{eq:hydro_eq})
with the density profile of $\rho_\mathrm{h}$.

The scattering-induced deceleration term is given by
\beqa
\ba_\mathrm{RPd} = -\bv_\mathrm{sub}\, \eta_\mathrm{d}\, \left(\frac{\sigma  |\bv_\mathrm{sub}|}{m}\right)\, \rho_\mathrm{h}, \label{eq:RP_deceleration}
\eeqa
where $\eta_\mathrm{d}$ is the deceleration fraction computed as \citep[see][]{2004ApJ...606..819M, 2018MNRAS.474..388K}
\beqa
\eta_\mathrm{d} &=& 1-4\int_{x/\sqrt{(1+x)^2}}^{1}\, \mathrm{d}y\, y^2\sqrt{y^2-x^2(1-y^2)}, \\
x &=& \frac{\bar{v}_\mathrm{esc,sub}}{\sqrt{|\bv|^2_\mathrm{sub}+\sigma^2_{v,\mathrm{h}}}}, \label{eq:x_for_eta}\\
\bar{v}_\mathrm{esc, sub} &=& \frac{1}{M_\mathrm{sub}}\int\, 4\pi r^2 \mathrm{d}r\, \rho_\mathrm{sub}\, \sqrt{\mathrm{-2\Phi_\mathrm{sub}}}.
\eeqa
In the above, $\Phi_\mathrm{sub}$ is the gravitational potential of the subhalo.
Note that we account for the bulk velocity of the subhalo as well as the random velocity of the particles inside the host halo in the computation of $\eta_\mathrm{d}$ \citep[see Appendix~A in][for details]{2018MNRAS.474..388K}.
Nevertheless, the effect of $\ba_\mathrm{RPd}$ is found to be almost negligible for our simulation results in this paper.

We solve Eq.~(\ref{eq:motion}) using a fourth-order Runge-Kutta method. 
It would be worth noting that we properly include the time evolution of the host halo density $\rho_\mathrm{h}$ across the subhalo orbit as in Subsection~\ref{subsec:gravothermal}.
To solve Eq.~(\ref{eq:motion}), we require a model of mass loss of the subhalo as well as the change of the subhalo density profile $\rho_\mathrm{sub}$ due to tidal effects and ram-pressure evaporation, described in the next Subsection.

\subsection{Mass loss}\label{subsec:mass_loss}

In the SIDM model, the infalling subhalo can lose its mass due to tidal stripping 
and ram-pressure evaporation effects. The former effect can predominantly remove mass from the outskirts of the subhalo, while the latter can affect the mass density in the entire region of the subhalo.

For the tidal stripping, we employ a commonly-used expression of the mass loss rate, given by
\beqa
\left(\frac{\mathrm{d}M_\mathrm{sub}}{\mathrm{d}t}\right)_\mathrm{TS} 
= -{\cal A} \frac{M_\mathrm{sub}(>r_t; t)}{q\tau_\mathrm{dyn}(R)}, \label{eq:mass_loss_TS}
\eeqa
where 
${\cal A}$ is a free parameter in the model,
$M_\mathrm{sub}(>r_t; t)$ represents the subhalo mass in the outskirts with $r>r_t$ at $t$,
$\tau_\mathrm{dyn}(R)$ is the dynamical time 
at the relative distance between the subhalo and the host centre being $R$,
and $q$ is a parameter with an order of unity.
To be specific, we define the dynamical time as
\beqa
\tau_\mathrm{dyn}(R) = \sqrt{\frac{\pi^2 R^3}{4 G M_\mathrm{h}(R)}},
\eeqa
and note that $M_\mathrm{h}(R)$ is the enclosed mass of the host and depends on time $t$.
We account for possible effects of (sub)halo concentrations at initial states by
setting $q = (c_\mathrm{sub}/c_\mathrm{h}/2)^{1/3}$ \citep[motivated by the results in][]{2021MNRAS.503.4075G}.
The value of ${\cal A}$ will be calibrated with our simulations.
The radius of $r_t$ is known as the tidal radius, and there are a number of different definitions \citep[e.g. see][for a brief overview]{2018MNRAS.474.3043V}.
In this paper, we adopt a phenomenological model of 
\beqa
r_t = \mathrm{min}(r_{t1}, r_{t2}),
\eeqa
with 
\beqa
\frac{r_{t1}}{R} &=& \left[\frac{M_\mathrm{sub}(r_{t1})/M_\mathrm{h}(R)}{2-(\mathrm{d}\ln M_\mathrm{h}/\mathrm{d}\ln r)_{r=R} + \left(v_\mathrm{tan, sub}/v_\mathrm{circ, h}(R)\right)^2}\right]^{1/3}, \label{eq:tidal_radius1}\\
\frac{r_{t2}}{R} &=& \left(\frac{M_\mathrm{sub}(r_{t2})}{M_\mathrm{h}(R)}\right)^{1/3}, \label{eq:tidal_radius2}
\eeqa
where $v_\mathrm{tan, sub} = |\bx_\mathrm{sub} \times \bv_\mathrm{sub}|/|\bx_\mathrm{sub}|$ 
is the instantaneous tangential velocity of the subhalo,
and $v_\mathrm{circ, h}(R) = \sqrt{GM_\mathrm{h}(R)/R^2}$ 
represents the circular velocity of a test particle in the host at the radius of $R$.
Note that one derives Eq.~(\ref{eq:tidal_radius1}) 
by assuming that the subhalo can be approximated as a point mass on a circular orbit \citep{1957ApJ...125..451V, 1962AJ.....67..471K}, while the assumption becomes invalid for more radial orbits.
Eq.~(\ref{eq:tidal_radius2}) has been proposed in \citet{1999ApJ...516..530K} to account for resonances between the gravitational force by the subhalo and the tidal force by the host \citep{1994AJ....108.1398W, 1994AJ....108.1403W, 1997ApJ...478..435W}.
If we can not find a non-trivial solution of $r_{t1}\neq0$ in Eq.~(\ref{eq:tidal_radius1}), 
we set $r_t = r_{t2}$.

For the ram-pressure evaporation, we adopt the mass loss rate below \citep[][]{2018MNRAS.474..388K}
\beqa
\left(\frac{\mathrm{d}M_\mathrm{sub}}{\mathrm{d}t}\right)_\mathrm{RPe} = -M_\mathrm{sub}\, \eta_\mathrm{e}\, \left(\frac{\sigma |\bv_\mathrm{sub}|}{m}\right)\, \rho_\mathrm{h}, \label{eq:mass_loss_RPe}
\eeqa
where $\eta_\mathrm{e}$ is the evaporation fraction computed as \citep[see][]{2004ApJ...606..819M, 2018MNRAS.474..388K}
\beqa
\eta_\mathrm{e} = \frac{1-x^2}{1+x^2},
\eeqa
and $x$ in the above is given by Eq.~(\ref{eq:x_for_eta}).

At each moment $t$, we can compute the mass loss of the subhalo during a small time interval of $\Delta t$ by using Eqs.~(\ref{eq:mass_loss_TS}) and (\ref{eq:mass_loss_RPe}). 
We then reset the subhalo mass of 
\beqa
M_\mathrm{sub} \rightarrow M_\mathrm{sub} +
\Delta t \left(\frac{\mathrm{d}M_\mathrm{sub}}{\mathrm{d}t}\right)_\mathrm{TS}+
\Delta t \left(\frac{\mathrm{d}M_\mathrm{sub}}{\mathrm{d}t}\right)_\mathrm{RPe},
\eeqa
and include effective tidal stripping effects on the subhalo density profile as
\beqa
\rho_\mathrm{sub}(r, t+\Delta t) = \rho_\mathrm{SIDM, \mathrm{sub}}(r, t+\Delta t) \, H(r; f_\mathrm{bound}, c_\mathrm{sub}),
\label{eq:density_stripped_sub}
\eeqa
where $\rho_\mathrm{SIDM, \mathrm{sub}}(r, t)$ is the model of Eq.~(\ref{eq:final_iso_density}) for the subhalo, $f_\mathrm{bound}$ is the bound mass defined as $M_\mathrm{sub}(t+\Delta t)/M_\mathrm{sub}(t=0)$,
and $H(r; f_\mathrm{bound}, c_\mathrm{sub})$ presents the change of the subhalo density 
profile due to the tidal stripping (referred to as the transfer function in the literature).
After updating the subhalo mass and its density profile, we then solve Eq.~(\ref{eq:motion}) to obtain the position and velocity of the subhalo at the time of $t+\Delta t$.
In practice, we set the time-step $\Delta t$ to be $10^{-4} T_r$ throughout this paper.

In Eq.~(\ref{eq:density_stripped_sub}), 
we assume that the ram-pressure effects are less important for the shape in the subhalo density profile,
but the tidal stripping plays a central role.
Tidal evolution of density profiles of infalling subhaloes has been investigated in \citet{2019MNRAS.485..189O, 2019MNRAS.490.2091G}
with a large set of N-body simulations of minor mergers for collision-less dark matter (i.e. $\sigma/m=0$).
\citet{2019MNRAS.490.2091G} has studied the tidal evolution of the subhalo density profile with respect to its initial counterpart and found that the structural evolution of a tidally truncated subhalo is predominantly determined by the bound mass fraction $f_\mathrm{bound}$ and the initial subhalo concentration.
We here adopt their calibrated model of the transfer function $H$ in Eq.~(\ref{eq:density_stripped_sub}).
The explicit form of $H$ is provided in Appendix~\ref{apdx:tranfer_func}.
It should be noted that \citet{2019MNRAS.490.2091G} calibrated the form of $H$ with the tidally stripped profile relative to the initial profile, but our model uses their transfer function for the time-evolving SIDM density profile. 
Although our model can reproduce the results in \citet{2019MNRAS.490.2091G} 
in the limit of $\sigma/m \rightarrow 0$ and $\rho_\mathrm{SIDM, sub} \rightarrow \rho_\mathrm{NFW, sub}$,
Eq.~(\ref{eq:density_stripped_sub}) should be validated with our N-body simulations for SIDM models.
We summarise our validation of Eq.~(\ref{eq:density_stripped_sub}) in Subsection~\ref{subsec:tidal_ev_SIDM}.

\begin{figure*}
 \includegraphics[width=2\columnwidth]{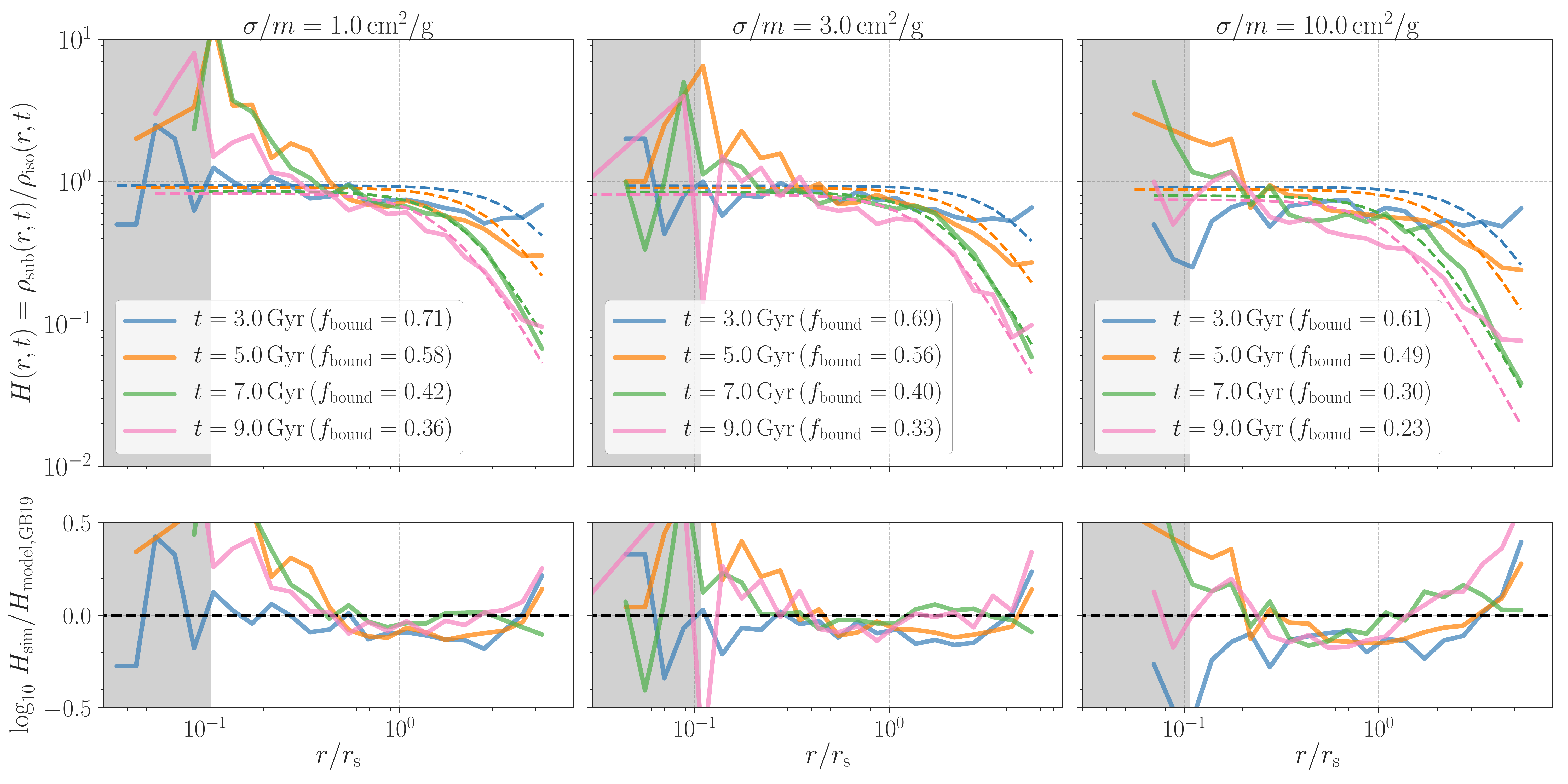}
 \caption{Structural evolution of density profiles of infalling subhaloes in SIDM models. From left to right, we show the results with the self-interacting cross section of $\sigma/m=1, 3$ and $10\, \mathrm{cm}^2/\mathrm{g}$, respectively.
 For each model, the upper panel shows the transfer function of the subhalo density profile (denoted as $H(r,t)$) measured in our N-body simulations. Different coloured lines represent the results at different epochs ($t=3, 5, 7$, and $9\, \mathrm{Gyr}$).
 The dashed lines in the upper panels are model predictions in \citet{2019MNRAS.490.2091G}. The lower panels summarises the fractional difference between the simulation results and the model predictions. 
 Note that numerical resolution effects would be important in the grey region in the figure.
 Although the model in \citet{2019MNRAS.490.2091G} has been calibrated with N-body simulations with $\sigma/m=0\, \mathrm{cm}^2/\mathrm{g}$,
 it can provide a reasonable fit to the simulation results with $1\le \sigma/m \,(\mathrm{cm}^2/\mathrm{g})\le 10$
 if the mass fraction of subhalo bound mass $f_\mathrm{bound}$ is set to the values in our N-body simulations. These results highlight that scattering processes between host- and sub-haloes are less important to determine the shape of the subhalo density profile, as long as we consider the cross section of $\sigma/m \simlt 10\, \mathrm{cm}^2/\mathrm{g}$.
 }
 \label{fig:test_GB19}
\end{figure*}

\subsection{For velocity-dependent cross sections}\label{subsec:treatment_vdependent_case}

We here explain how our model can be applied for velocity-dependent cross sections $\sigma(v)/m$.
Suppose that we solve the time evolution of 
the system with an time interval of $\Delta t$.
At the $n$-th time-step $t=t_n$, 
our model follows procedures below;

\begin{enumerate}
\item We first determine the time evolution of density profiles for isolated host- and sub-haloes as in Subsection~\ref{subsec:gravothermal}.
For this purpose, we set effective cross sections to
\beqa
\left(\frac{\sigma}{m}\right)_\mathrm{eff}
&\equiv& \frac{\langle \sigma v/m \rangle}{\langle v \rangle}, \label{eq:sigma_m_eff}
\\
\langle \sigma v /m \rangle &=& 
\int_0^{\infty} \mathrm{d}v\, v\, \sigma(v)/m\, f(v; v_c), \\
\langle v \rangle &=&
\int_0^{\infty} \mathrm{d}v\, v\, f(v;v_c),
\eeqa
where $f(v;v_c)$ represents the distribution function of relative velocity of particles in the host or subhalo,
and $v_c$ determines a typical velocity scale.
We define Eq.~(\ref{eq:sigma_m_eff}) with a velocity-weighted quantity because the number of particles scattered per unit time ($\propto \langle \sigma v/m \rangle$) is expected to be relevant to the evolution of SIDM density profiles.
In this paper, we assume $f(v;v_c)$ as a Maxwell-Boltzmann distribution 
for relative velocities;
\beqa
f(v;v_c) = \frac{4 v^2_c \exp(-v^2/v^2_c)}{\sqrt{\pi}v^3_c}, \label{eq:MB_dist}
\eeqa
providing that $\langle v \rangle = v_c$.
For a given halo/subhalo density profile at $t=t_{n-1}$, 
we determine the 1D velocity dispersion $\sigma_v(r)$ by Eq.~(\ref{eq:hydro_eq}) and set $v_c = 4 \sigma_v(r_s)/\sqrt{\pi}$ where $r_s$ is 
the scaled radius at the initial NFW profile.
We then take the corresponding SIDM density profile at $\sigma/m = (\sigma/m)_\mathrm{eff}$ and the moment of $t=t_n$ from a pre-stored table of $\rho_\mathrm{SIDM}$ given by Eq.~(\ref{eq:final_iso_density}) for $v$-independent cross sections.

\hspace{12pt}

\item We then solve the equation of motion of the subhalo as in Subsection~\ref{subsec:eq_motion}.
To determine the ram-pressure deceleration term of Eq.~(\ref{eq:RP_deceleration}), we substitute $\sigma/m$
for $\sigma(|\bv_\mathrm{sub, n-1}|)/m$,
where $\bv_\mathrm{sub, n-1}$ is the bulk velocity of the subhalo at $t=t_{n-1}$.
Using the time-step of $\Delta t$, 
we also set the mass loss of the infalling subhalo and update the shape of the subhalo density profile as in Eq.~(\ref{eq:density_stripped_sub}).
For the velocity-dependent cross section, we compute the mass loss of Eq.~(\ref{eq:mass_loss_RPe}) by setting $\sigma/m = \sigma(|\bv_\mathrm{sub, n-1}|)/m$.

\hspace{12pt}

\item After updating the bound mass, position, velocity, and the density profile of the subhalo, we go back to the step (i) to determine the density profiles at $t=t_{n+1}$.

\end{enumerate}

\section{Results}\label{sec:results}

This section presents main results in our paper.
Those include the structural evolution of subhalo density profiles with dark matter self-interactions, detailed comparisons with our semi-analytic model and the simulation outputs,
and discussion on differences between our model 
and others in the literature.

\subsection{Structural evolution of SIDM subhaloes}\label{subsec:tidal_ev_SIDM}

We first study density profiles of infalling SIDM subhaloes at different epochs. As the subhalo orbit is evolved, the density profile is modified by gravitational interactions as well as the self-interaction of dark matter particles in the host and subhalo.

For ease of comparison, we run N-body simulations of 
an isolated halo with its initial mass of $10^{9}M_\odot$ and concentration of $6$, but varying $\sigma/m=1, 3$, and $10\, \mathrm{cm}^2/\mathrm{g}$. 
These isolated haloes are evolved by $10\, \mathrm{Gyr}$ with a snapshot interval of $0.1\, \mathrm{Gyr}$.
We then characterise the structural evolution of infalling subhalo density profiles as
\beqa
H(r,t) \equiv \frac{\rho_\mathrm{sub}(r,t)}{\rho_\mathrm{iso}(r,t)},
\eeqa
where $\rho_\mathrm{sub}(r,t)$ is the density profile of infalling subhaloes, and $\rho_\mathrm{iso}(r,t)$ represents the counterpart for isolated haloes with the same initial density profiles as the subhaloes.

Figure~\ref{fig:test_GB19} summarises 
our measurements of $H(r,t)$.
At each column, upper and lower panels present the results at $\sigma/m=1, 3$ and $10\, \mathrm{cm}^2/\mathrm{g}$
from left to right.
Solid lines in the upper panel show the function of $H(r,t)$
in our simulations and the colour difference indicates the difference in the epoch $t$.
The coloured dashed lines in the upper panel are the prediction in \citet{2019MNRAS.490.2091G} with the simulated value of the bound mass fraction $f_\mathrm{bound}$.
The fractional difference between the simulation results and the model prediction is shown in the lower panels.

\begin{figure*}
 \includegraphics[width=2\columnwidth]{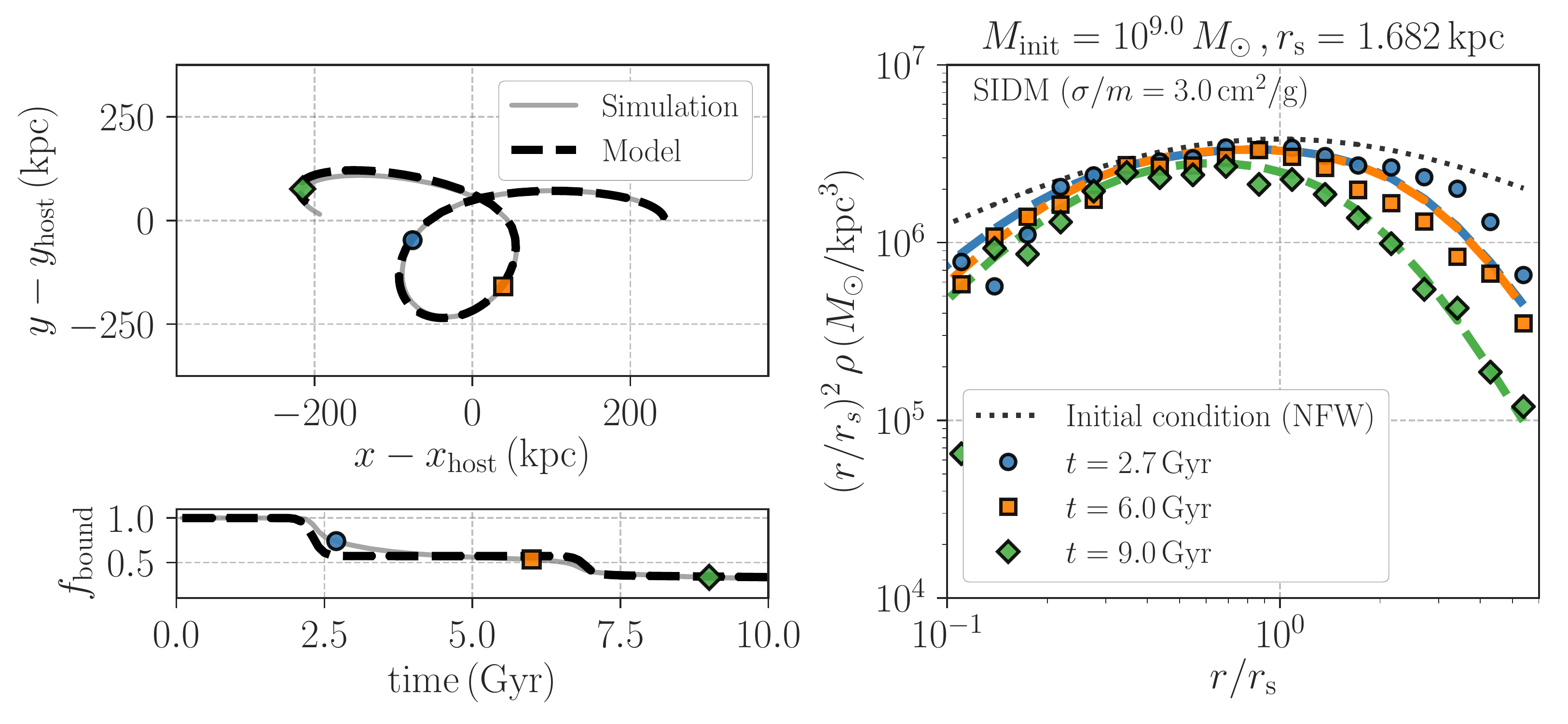}
 \caption{Comparisons with N-body simulation results and our semi-analytic model of infalling subhaloes. In this figure, we assume a velocity-independent cross section of $\sigma/m=3\, \mathrm{cm}^2/\mathrm{g}$. The top left panel shows the orbital evolution of the subhalo over 10 Gyr,
 while the bottom left presents the mass evolution.
 The right panel summarises the time evolution of the subhalo density profile.
 In the right, blue circles, orange squares, and green diamonds 
 represent the simulation results at $t=2.7$, 6.0, and 9.0 Gyr, respectively.
 In each panel, dashed lines are the model predictions.
 }
 \label{fig:model_vs_sim_sigm3}
\end{figure*}

We find that the structural evolution of SIDM subhaloes can be approximated as the model in \citet{2019MNRAS.490.2091G}, even though the model has been calibrated with 
the collision-less N-body simulations.
As long as the cross section is set to smaller than $\sim10\, \mathrm{cm^2}/\mathrm{g}$, the ram-pressure evaporation is less important to set the shape of the subhalo density profile.
We here note that a reasonable match between the simulation results and the model in \citet{2019MNRAS.490.2091G} occurs only when we use the value of $f_\mathrm{bound}$ in the simulations.
This highlights that a precise model of the mass loss is important to determine the density profile of the subhalo at outskirts across its orbit. 
Also, the results in Figure~\ref{fig:test_GB19} support that 
our approximation of Eq.~(\ref{eq:density_stripped_sub}) would be valid if we can predict the density profile of 
SIDM halos in isolation.
More detailed comparisons with the simulation results and Eq.~(\ref{eq:density_stripped_sub}) are presented in the next Subsection.

\subsection{Comparison with simulation results and model predictions}

We here summarise comparisons with our N-body simulation results and model predictions as in Section~\ref{sec:model}.

\begin{figure*}
 \includegraphics[width=2\columnwidth]{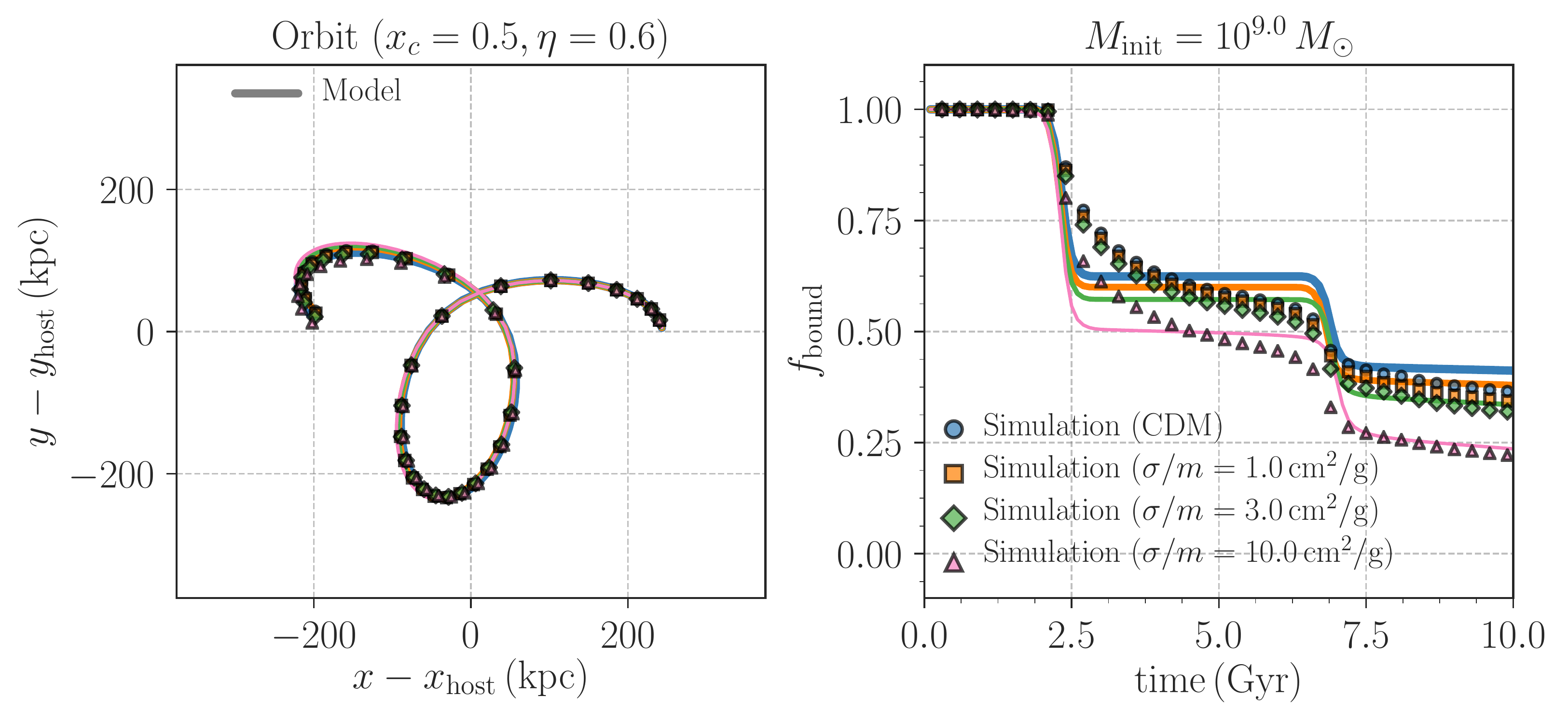}
 \caption{The orbital and mass evolution of an infalling subhalo with its initial mass of $10^{9}\, M_\odot$ as a function of $\sigma/m$. In this figure, $\sigma/m$ is assumed to be velocity-independent. In each panel, the blue circles, orange squares, green diamonds, and pink triangles represent the simulation results at $\sigma/m=0, 1, 3$,
 and $10\, \mathrm{cm}^2/\mathrm{g}$, respectively. Our model predictions are shown 
 by different lines, providing a reasonable fit to the simulation results.
 }
 \label{fig:model_vs_sim_various_sigm}
\end{figure*}

\subsubsection{Varying cross sections}

We first investigate the dynamical evolution of infalling subhaloes with their initial mass of $10^{9} M_\odot$ and a fixed subhalo orbital parameter 
as a function of the self-interaction cross section $\sigma/m$.
For this purpose, we use the fiducial simulation runs of CDM, SIDM1, and SIDM3, and SIDM10 in Table~\ref{tab:sim_params}.

Figure~\ref{fig:model_vs_sim_sigm3} summarises the simulation outputs of the infalling subhalo for the SIDM3 run ($\sigma/m=3\, \mathrm{cm}^2/\mathrm{g}$) as well as our model predictions.
In the left panels, grey lines represent the simulation results, while the dashed lines 
are our model predictions. 
For this figure, we set a parameter for the mass loss (see Eq.~\ref{eq:mass_loss_TS}) 
to ${\cal A}=0.65$. 
Our model provides an accurate fit to the subhalo orbit in our simulation over 10 Gyr,
and the overall evolution of the subhalo mass can be captured 
by the simple model in Subsection~\ref{subsec:mass_loss}.
In the right panel, we compare the subhalo density profile at different epochs.
The simulation results are shown by coloured symbols, and the dashed lines show the model predictions.
The figure demonstrates that the structural evolution of the subhalo density profile
can be explained by our phenomenological model of Eq.~(\ref{eq:density_stripped_sub}).
The time evolution at $r\simlt r_s$ can be well determined by the gravothermal fluid model 
with a correction (see Eq.~\ref{eq:final_iso_density}), 
while the density at outskirts ($r \simgt r_s$) is suppressed mostly by 
tidal stripping processes.

\begin{figure*}
 \includegraphics[width=2\columnwidth]{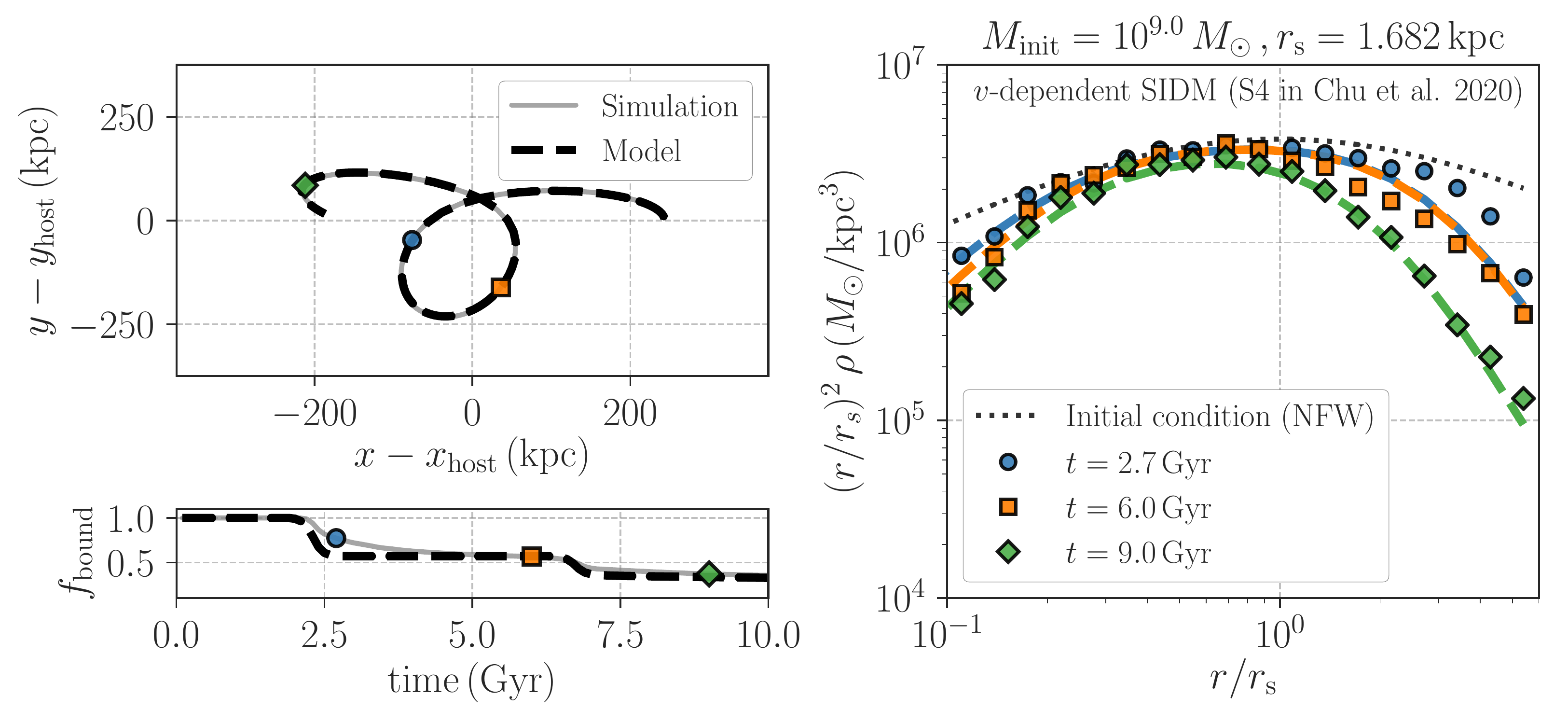}
 \caption{Similar to Figure~\ref{fig:model_vs_sim_sigm3}, but we consider a velocity-dependent cross section given by Eq~(\ref{eq:vdep_sigma_over_m}).
 }
 \label{fig:model_vs_sim_vdepend_sigm}
\end{figure*}

Figure~\ref{fig:model_vs_sim_various_sigm} shows how the dynamical evolution of the subhalo can depend on the cross section $\sigma/m$. The orbital evolution of the subhalo with different $\sigma/m$
are summarised in the left, while the right shows the evolution of the subhalo mass over 10 Gyr.
In each panel, solid lines represent our model predictions, providing a reasonable fit to the simulation results for various cross sections.
We find that the model works when the parameter $\cal A$ is set to $0.55$, $0.60$, $0.65$ and $0.75$
for the simulations with $\sigma/m = 0, 1, 3$, and $10\, \mathrm{cm}^2/\mathrm{g}$, respectively.
This marginal $\sigma/m$-dependence of the model parameter $\cal A$
can be important in practice, 
especially when one would constrain the SIDM by using observations of MW satellites.
We also note that the subhalo mass is more suppressed as $\sigma/m$ becomes larger in our simulations
and this looks compatible with recent studies \citep[e.g.][]{2020PhRvL.124n1102S}.

We then examine the velocity-dependent model of $\sigma/m$ as in Eq.~(\ref{eq:vdep_sigma_over_m})
by using the vSIDM run (see Table~\ref{tab:sim_params}).
Figure~\ref{fig:model_vs_sim_vdepend_sigm} summarises the comparison of the simulation results with
our semi-analytic model.
Note that we set ${\cal A} = 0.65$ in Figure~\ref{fig:model_vs_sim_vdepend_sigm}.
The figure highlights that our treatment in 
Subsection~\ref{subsec:treatment_vdependent_case} can explain the simulation results
with an appropriate choice of ${\cal A}$.

\subsubsection{Varying subhalo orbits}

\begin{figure*}
 \includegraphics[width=2\columnwidth]{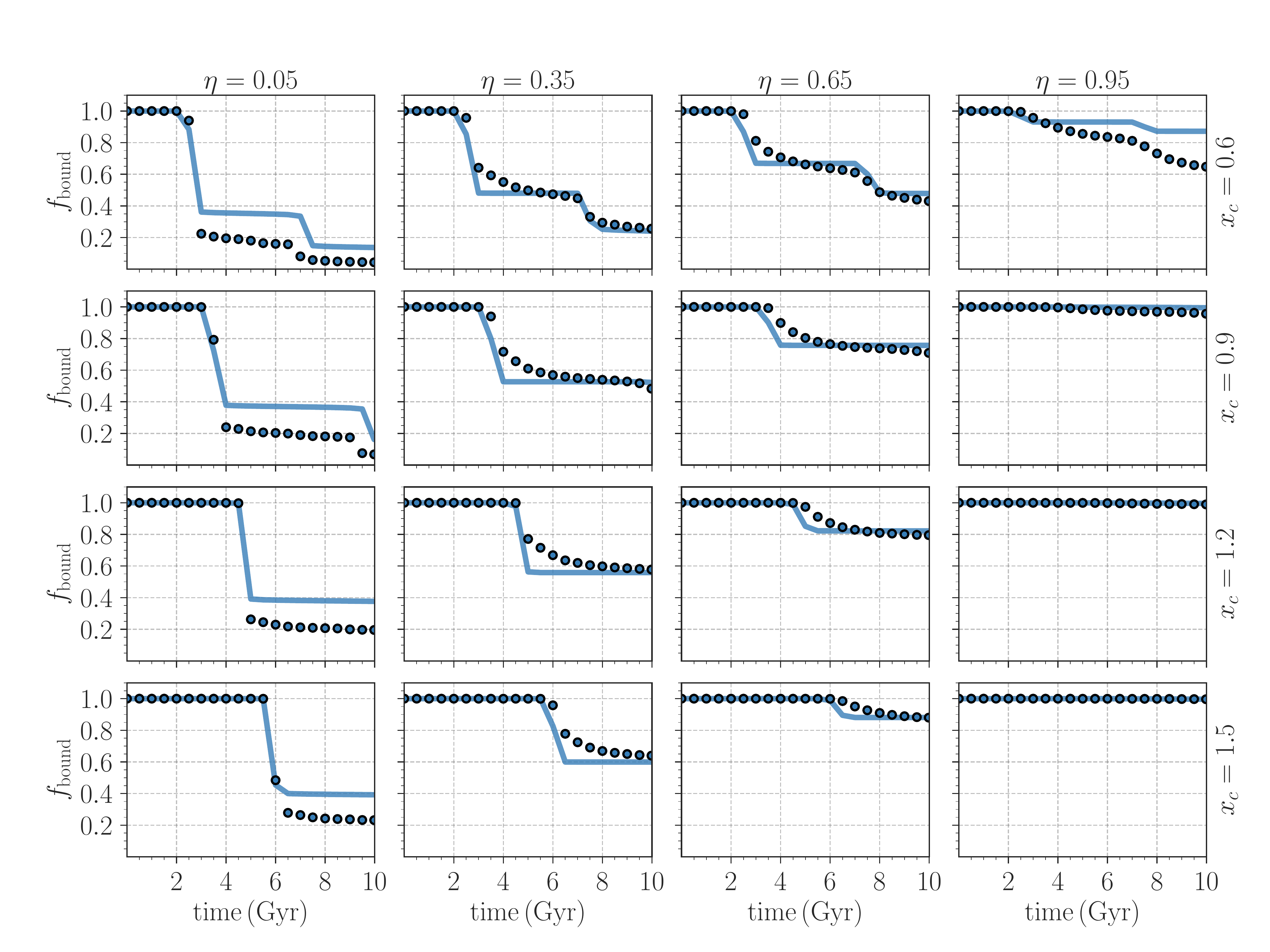}
 \caption{The mass evolution of infalling subhaloes at various orbits for the SIDM model
 with the velocity-independent cross section of $1\, \mathrm{cm}^2/\mathrm{g}$.
 In each panel, the blue circles show the simulation results, while the line presents our model prediction. The subhalo orbits are characterised by two parameters of $x_c$ and $\eta$.
 The results with $\eta=0.05, 0.35, 0.65$, and $0.95$ are shown from left to right, while
 we increase $x_c$ as $x_c=0.6, 0.9, 1.2$, and $1.5$ from top to bottom. Note that larger $x_c$ corresponds to longer orbital period, and smaller $\eta$ provides more radial orbits (see Subsection~\ref{subsec:inital_condition} for details).
 }
 \label{fig:model_vs_sim_mass_loss_various_orbit}
\end{figure*}

We next study the impact of subhalo orbits on the subhalo mass loss in SIDM models.
We examine 16 different sets of our orbital parameters $(x_c, \eta)$ as in Table~\ref{tab:sim_params}
assuming the velocity-independent cross section of $\sigma/m=1\, \mathrm{cm}^2/\mathrm{g}$.

Figure~\ref{fig:model_vs_sim_mass_loss_various_orbit} summarises the time evolution of infalling subhalo masses as a function of $(x_c, \eta)$. The blue circles in the figure represent the simulation results, while the solid lines show our model predictions.
We assume ${\cal A} = 0.60$ for every model prediction in the figure.
We find that our model can provide a reasonable fit to the simulation results with $\eta \simgt 0.35$
and a range of $0.6 \le x_c \le 1.5$, but a sizeable difference between the simulation results and our model can be found at an extreme value of $\eta \simeq 0.05$.
Note that orbits with $\eta \simlt 0.2$ rarely happens in cosmological simulations of collision-less dark matter \citep[e.g.][]{2015MNRAS.448.1674J}.
Even for the orbits at $\eta=0.05$, our model can explain overall trends in the time evolution of the subhalo mass with a level of $20-30\%$.

\subsubsection{Model precision of subhalo density profiles}

\begin{figure*}
 \includegraphics[width=2\columnwidth]{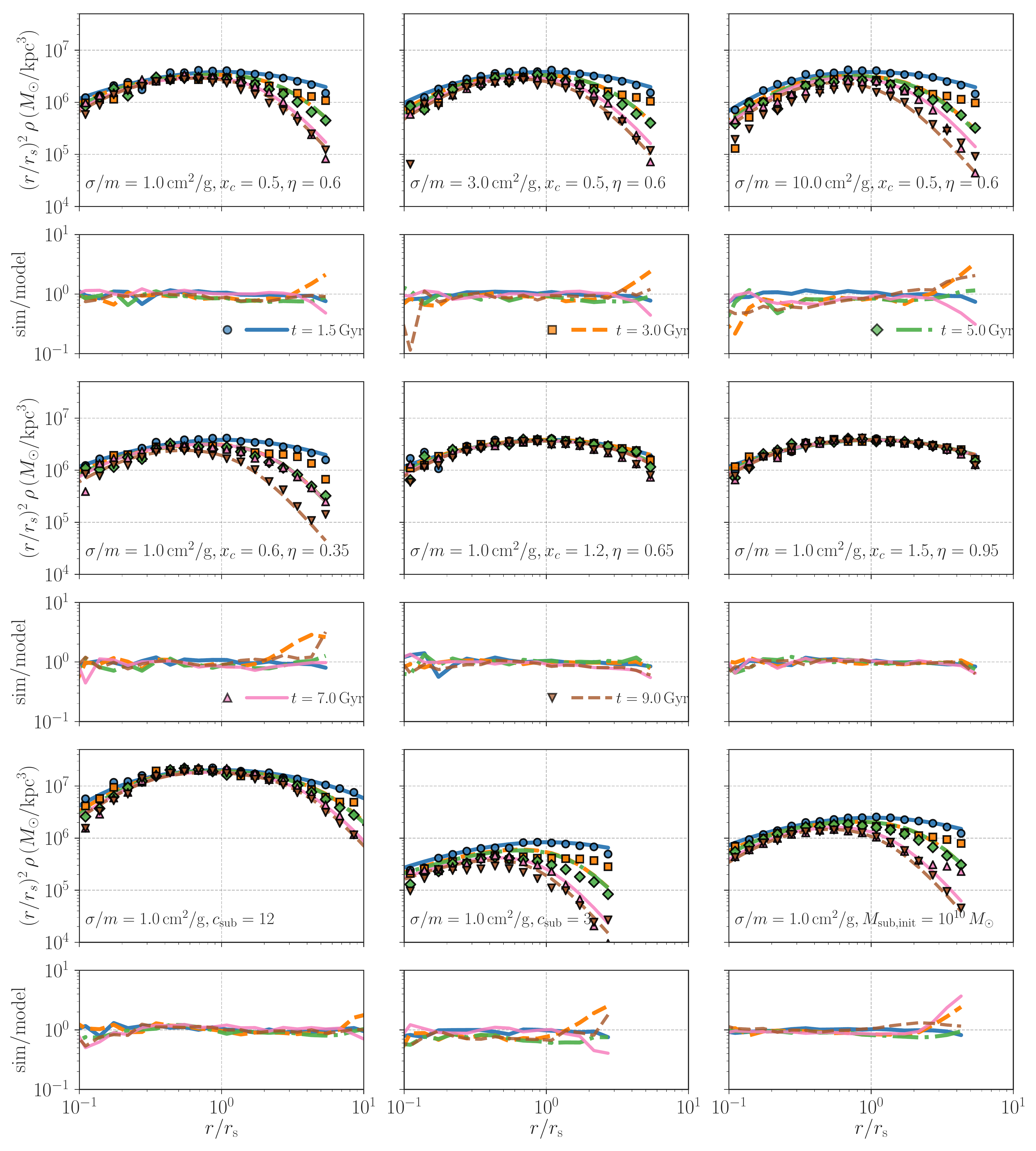}
 \caption{Precision of the subhalo density profile by our semi-analytic model.
 At the first, third, and fifth rows, coloured symbols show the density profiles in our N-body simulations at different epochs of $t=1.5, 3, 5, 7$ and 9 Gyr, while the coloured lines are our model predictions. Each panel at the second, fourth, and six rows show the ratio between the density profile in our N-body simulations and our model counterparts for the ease of comparison.
 At the first and second rows, we show the results at a fixed subhalo orbit, but increase the cross section as $\sigma/m=1, 3$, and $10\, \mathrm{cm}^2/\mathrm{g}$ from left to right.
 At the third and fourth rows, we fix the cross section to $\sigma/m=1\, \mathrm{cm}^2/\mathrm{g}$, but change the subhalo orbits. At the fifth and sixth rows, we examine different density profiles of subhaloes at $t=0$ in the SIDM model with $\sigma/m=1\, \mathrm{cm}^2/\mathrm{g}$.
 }
 \label{fig:model_precision_sat_profile}
\end{figure*}

We then investigate the subhalo density profiles at various initial conditions as well as examine the dependence on the self-interaction cross section $\sigma/m$.
Figure~\ref{fig:model_precision_sat_profile} compares the subhalo density profiles in our N-body simulations with the model counterparts.
In this figure, the first, third and fifth rows summarise the subhalo density profiles in various simulation runs.
At those rows, different coloured symbols represent the subhalo density profiles in the simulation at different epochs of $t=1.5, 3, 5, 7$ and $9\, \mathrm{Gyr}$, while the coloured lines are the counterparts by our model prediction.
At the second, fourth, and sixth rows, individual panels show the ratio between the simulation results and our model predictions for comparison.

At the first and second rows, we show the results as varying $\sigma/m$ for a fixed initial condition of the subhalo. We observe that our model can reproduce the subhalo density profiles in the simulations
with a level of $\sim0.1$ dex in a range of $r/r_s \simgt 1$ when varied the cross section $\sigma/m$. The model precision becomes worse as we increase $\sigma/m$, implying that effects of gravothermal instability may be required to be revised for a better model.

Three panels at the third and fourth rows summarise the comparisons at different orbital parameters $(x_c, \eta)$ for the SIDM model with $\sigma/m = 1\, \mathrm{cm}^2/\mathrm{g}$.
As long as the orbital parameter is set to be $\eta \simgt 0.35$,
our model can provide an accurate fit to the simulation results. 
Note that a small value of $\eta$ corresponds to a highly elongated orbit around 
the host.
When setting an extreme condition of $\eta =0.05$, we observed that our model precision gets worse (but the model has a 0.5 dex level precision). For tidal effects, our model partly relies on the assumption of the subhalo on a circular orbit (Eq.~\ref{eq:tidal_radius1}). Hence the model would tend to be invalid for more radial orbits.

In the panels at the fifth and sixth rows, we can see the effect of initial conditions of subhaloes for the SIDM model with $\sigma/m = 1\, \mathrm{cm}^2/\mathrm{g}$.
The left panel at the fifth row shows the comparisons when we assume an initial subhalo density profile with a higher concentration, while the middle bottom panel presents the results with the subhalo with a lower concentration at $t=0$. 
We find that our model can reproduce the simulation results with a level of $\sim0.2$ dex 
for a wide range of the subhalo concentration at their initial density.
The model precision gets worse for the lower-concentration subhalo, 
indicating that a more detailed calibration of the gravothermal fluid model (see Eq.~\ref{eq:final_iso_density}) and the tidal stripping model (see Eq.~\ref{eq:density_stripped_sub}) are beneficial. 
The right panel at the fifth row in the figure summarises the comparisons when we increase the subhalo mass at its initial state as $M_\mathrm{sub}=10^{10}\, M_\odot$. 
We do not observe any systematic trends in the difference between the simulation results and our model predictions even if we increase the initial subhalo mass.

\begin{figure*}
 \includegraphics[width=2\columnwidth]{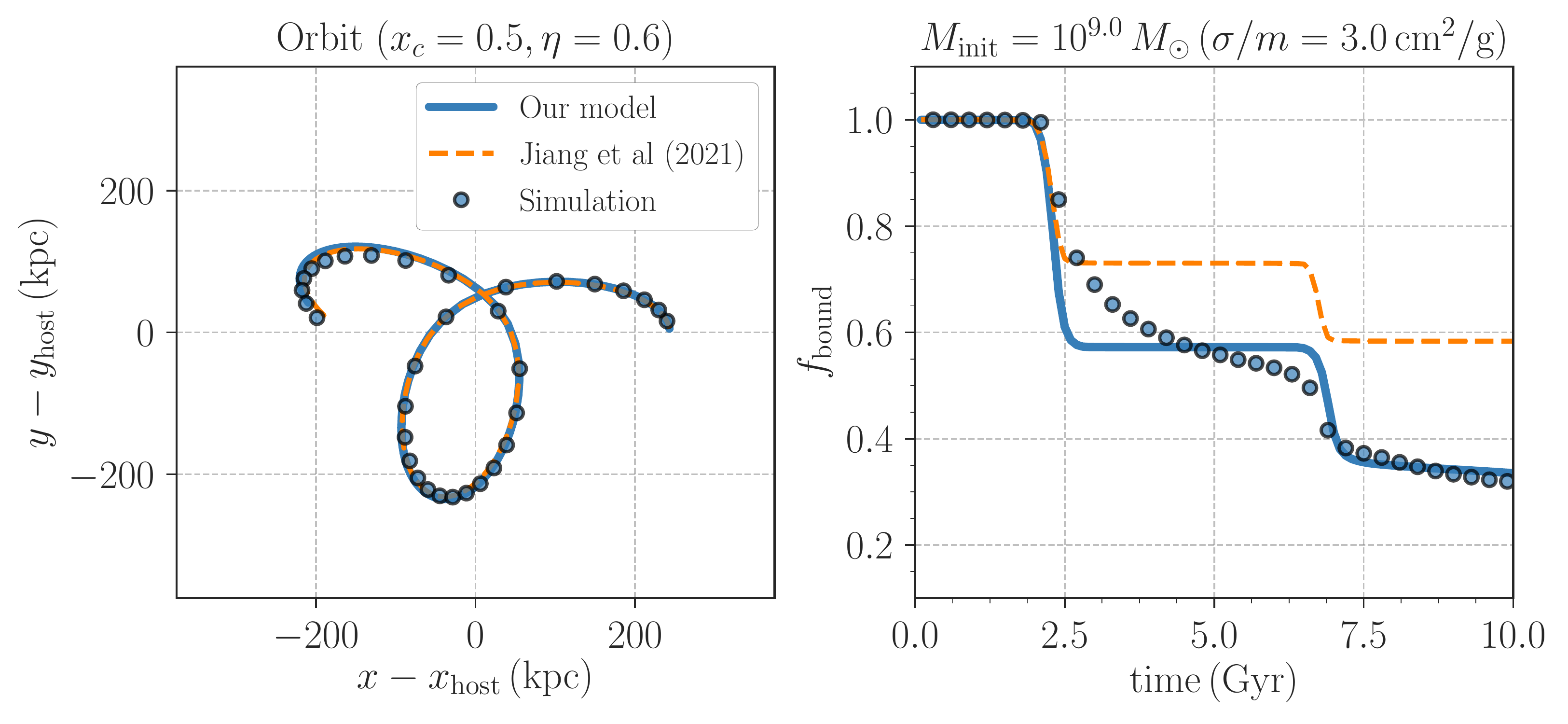}
 \caption{Similar to Figure~\ref{fig:model_vs_sim_various_sigm},
 but we include the model prediction in \citet{2021arXiv210803243J}.
 In this figure, we assume a velocity-independent cross section of $\sigma/m=3\, \mathrm{cm}^2/\mathrm{g}$.
 The blue circles show our simulation results, the solid lines are our model predictions, and the orange dashed lines represent the model in \citet{2021arXiv210803243J}. 
 }
 \label{fig:model_comparison}
\end{figure*}

\subsection{Comparison with previous studies}

In the aforementioned sections, we introduced a semi-analytic model of infalling subhaloes 
and made detailed comparisons with ideal N-body simulation results and the model predictions.
We here discuss differences among our model and others in the literature.

\subsubsection{Time evolution of density profiles of single SIDM haloes}

Our model assumes a gravothermal fluid model based on the calibration of the thermal conductivity $\kappa$ in Eq.~(\ref{eq:energy_flux}) in \citet{2011MNRAS.415.1125K}, 
whereas we further include a correction based on our N-body simulations of isolated SIDM haloes as in Eq.~(\ref{eq:final_iso_density}).
Previous studies have reported different models of $\kappa$ for isolated and cosmological N-body simulations \citep[e.g.][]{2002ApJ...568..475B, 2011MNRAS.415.1125K, 2019PhRvL.123l1102E, 2020PhRvD.101f3009N}. 
Also, the hydrostatic equilibrium (Eq.~\ref{eq:hydro_eq}) is not always valid in SIDM haloes
at small cross sections \citep[e.g.][]{2020PhRvD.101f3009N}.
Hence, a correction of the gravothermal fluid model would be needed for a precise 
modelling of time evolution of SIDM density profiles.
Nevertheless, it would be worth noting that we correct the gravothermal fluid model with a level of 10-50\% over 10 Gyr. From a qualitative point of view, the fluid model in \citet{2011MNRAS.415.1125K} provides a fit to our N-body simulation results.

For another approach, \citet{2021MNRAS.501.4610R} 
introduced a mapping method from a given NFW profile to
an isothermal density profile based on Jeans equations, referred to as isothermal Jeans modelling.
In the isothermal Jeans modelling, one assumes that a SIDM halo follows an isothermal density profile at the radius smaller than $r_c$, while the NFW profile remains unchanged at outer radii.
The isothermal Jeans modelling is found to be valid when one predicts the density profile of a SIDM halo at a given epoch, but a proper choice of $r_c$ is required 
to explain simulation results on a case-by-case basis.
Hence, the isothermal Jeans modelling is less relevant to predicting the time evolution of the SIDM density profile.

\subsubsection{Evolution of infalling subhaloes}

Our model assumes that the motion of infalling SIDM subhaloes is governed by 
Eq.~(\ref{eq:motion}) as same as in \citet{2021arXiv210803243J}.
The model of \citet{2021arXiv210803243J}, referred to as J21 model, 
assumes that 
(i) an isolated SIDM halo follows a cored profile with a characteristic core radius 
where every particle is expected to have interacted once within a time,
(ii) a parameter of the mass loss in Eq.~(\ref{eq:mass_loss_TS}) is fixed to ${\cal A}=0.55$ 
as expected in the collision-less dark matter \citep{2021MNRAS.503.4075G}, 
and 
(iii) the mass loss by tidal stripping effects (Eq.~\ref{eq:mass_loss_TS}) truncates the subhalo boundary radius and the mass loss by self-interactions (Eq.~\ref{eq:mass_loss_RPe}) decreases the amplitude in the subhalo density.
We also refer the readers to a brief description of the J21 model in Appendix~\ref{apdx:model_J21}.

Figure~\ref{fig:model_comparison} summarises the comparison with the J21 model and ours 
for the SIDM with the cross section of $\sigma/m = 3\, \mathrm{cm}^2/\mathrm{g}$.
We find that the difference in the subhalo orbit is very small.
On the time evolution of the subhalo mass, an appropriate choice of the parameter ${\cal A}$
is needed to provide a better fit to our simulation results.
Note that \citet{2021arXiv210803243J} assumes 
a static NFW gravitational potential for the host halo in their analysis.
Hence, the orbital evolution of infalling subhaloes in \citet{2021arXiv210803243J}
may be less affected by choices of the model, whereas 
the J21 model would have a $50\%$-level uncertainty in predicting the time evolution of the subhalo mass over $\sim10\, \mathrm{Gyr}$. 

Recently, \citet{2021MNRAS.503..920C} has developed a semi-analytic model of infalling subhaloes 
in a static host based on a gravothermal fluid model and 
derived an interesting constraint of SIDM models with  observations of MW dwarf spheroidal galaxies.
The model in \citet{2021MNRAS.503..920C} incorporated the gravothermal fluid model 
with the tidal evolution of subhaloes \citep{2018MNRAS.474.3043V, 2019MNRAS.490.2091G}, 
accounting for the gravothermal collapse effects accelerated by the tidal stripping \citep{2020PhRvD.101f3009N, 2020PhRvL.124n1102S}.
However, the model computes the mass loss rate assuming a circular subhalo orbit
and does not include the mass loss by the self-scattering-induced evaporation.
This simplification can affect the subhalo mass at each moment.
Because the gravothermal instability depends on how the subhalo mass density is tidally stripped,
further developments would be interesting for a precise modelling 
of the gravothermal collapse effects in infalling subhaloes.
Note that our model ignores the gravothermal instability induced by tidal stripping effects, 
while it can solve the orbital and structural evolution of subhaloes in a self-consistent way.

\section{Limitations}\label{sec:limitations}

Before concluding, we summarise the major limitations 
in our semi-analytic model of infalling subhaloes in a MW-sized host halo. 
The following issues will be addressed in future studies.

\subsection{Baryonic effects}

In this paper, we do not consider any baryonic effects.
Baryons can affect our semi-analytic model in various ways.

The presence of stellar and gas components is common in most of real galaxies.
The baryons at the galaxy centre can deepen the gravitational potential compared 
to dark-matter-only predictions.
This allows an effective temperature of SIDM particles to have a flat 
or negative gradient in the radius, 
leading to decrease the size of SIDM core as well as increase the central SIDM density
in baryon-dominated galaxies \citep{2014PhRvL.113b1302K, 2017PhRvL.119k1102K}.
These back-reaction effects between baryons and SIDM have been investigated in isolated N-body simulations \citep{2018MNRAS.479..359S} and cosmological zoom-in simulations \citep{2014MNRAS.444.3684V, 2019MNRAS.490..962F, 2019MNRAS.490.2117R, 2021MNRAS.507..720S}.
Interestingly, the simulations in \citet{2018MNRAS.479..359S} showed 
that the SIDM core in a MW-sized halo can expand at early phases and contract later.
This time variation can be important to predict orbits of infalling subhaloes 
in a realistic MW-sized galaxy.

In addition, the presence of stellar disc at the host centres can severely affect
the mass loss of infalling subhaloes.
\citet{2010ApJ...709.1138D} showed that subhaloes in the inner regions of the halo are 
efficiently destroyed in the presence of time-evolving stellar disc components, 
while \citet{2017MNRAS.471.1709G} found that this suppression in 
the subhalo abundance can be explained by adding an embedded central disc potential 
to dark-matter-only simulations.
Isolated N-body simulations also play important roles in studying the depletion of subhaloes 
in details \citep[e.g][]{2010MNRAS.406.1290P, 2017MNRAS.465L..59E}.
Recently, \citet{2022MNRAS.509.2624G} have explored the impact of a galactic disc potential 
on the subhalo populations in MW-like haloes with their semi-analytic modelling. 
We expect that our semi-analytic model can be useful to investigate the effects of 
stellar disc components in the SIDM model 
by adding a stellar disc potential in the equation of motion (Eq.~\ref{eq:motion}).

\subsection{Gravothermal collapse}

The gravothermal instability induces dynamical collapse of the SIDM core.
This effect can be partly taken into account in our semi-analytic model
with the gravothermal fluid model (see Subsection~\ref{subsec:gravothermal}).
Note that the gravothermal fluid model of isolated SIDM haloes 
predicts the core collapse over time, but it rarely happens within a Hubble time \citep[e.g.][]{2002ApJ...568..475B}.
Our model still assumes that the gravothermal collapse occurs regardless of the tidal stripping effects, but this is not the case for some specific conditions \citep{2020PhRvD.101f3009N, 2020PhRvL.124n1102S}.
\citet{2020PhRvD.101f3009N} found that the core collapse in the SIDM density can realise within a Hubble time for $\sigma/m \simlt 10\, \mathrm{cm}^2/\mathrm{g}$ if the initial subhalo density is significantly truncated, while \citet{2020PhRvL.124n1102S} showed that the evolution of the SIDM core 
is sensitive to the concentration in the initial subhalo density.
Motivated by those findings, \citet{2021MNRAS.503..920C} developed a gravothermal fluid model 
of tidally stripped subhaloes with focus on a large self-interacting cross section of $20-150\, \mathrm{cm}^2/\mathrm{g}$.
A calibration of the gravothermal fluid model in \citet{2021MNRAS.503..920C}
with N-body simulations would be an interesting direction of future studies.

\subsection{Comparisons with cosmological simulations}

Our semi-analytic model has been calibrated with isolated N-body simulations.
This indicates that our results may be affected by cosmological environments at the outermost radii.
A lumpy and continuous mass accretion in an expanding universe can heat SIDM haloes, slowing the gravothermal core collapse.
Detailed comparisons with our gravothermal fluid model of Eq.~(\ref{eq:final_iso_density})
with cosmological SIDM N-body simulations \citep[e.g.][]{2013MNRAS.430...81R, 2015MNRAS.453...29E} 
can reveal how important environmental effects are in predicting time evolution of the SIDM density profiles.

The evolution of infalling subhaloes can be affected by other floating subhaloes in the host.
The subhaloes should gravitationally interact with each other, and induce perturbations in the host gravitational potential.
These complex effects might affect the orbital and structural evolution of infalling subhaloes.
To examine these, it would be worth comparing our semi-analytic model with zoom-in simulation results of MW-sized cosmological haloes \citep[e.g.][]{2022PhRvD.105b3016E}.

\section{Conclusions and discussions}\label{sec:conclusions}

In this paper, we have studied the evolution of a $10^{9}\,M_\odot$ 
subhalo infalling onto a MW-sized host halo
in the presence of self-interactions among dark matter particles.
We have performed a set of dark-matter-only N-body simulations of halo-subhalo minor mergers by varying self-interacting cross sections $\sigma/m$,
subhalo orbits, and initial conditions of subhalo density profiles.
For comparisons, we developed a semi-analytic model of infalling subhaloes 
in a given host halo by combining a gravothermal fluid model with subhalo mass losses
due to tidal stripping and ram-pressure-induced effects.
We then made detailed comparisons with our simulation results and the semi-analytic model,
allowing to improve physical understanding of self-interacting dark matter (SIDM) substructures.
Although our study imposes several assumptions, we gained meaningful insights as follows:

\begin{enumerate}

\item In our N-body simulations for a range of $\sigma/m \simlt 30\, \mathrm{cm}^2/\mathrm{g}$, the fluid model with the thermal conductivity calibrated in \citet{2011MNRAS.415.1125K} can qualitatively explain the time evolution of the SIDM core in an isolated halo whose initial density follows a NFW profile, but we also observe systematic differences between the simulation results and the fluid model over 10 Gyr. We provided a simple correction of the model as in Eq.~(\ref{eq:final_iso_density}). 
Our corrected gravothermal fluid model allows to predict the time evolution of SIDM density profiles over 10 Gyr with a $10\%$-level precision.

\vspace{6pt}

\item The structural evolution of infalling subhaloes can be explained by the prediction for collision-less dark matter as proposed in \citet{2019MNRAS.490.2091G}, even if we include the self-interaction of dark matter particles. The evaporation due to self-interacting ram pressure can not alter the SIDM density profile in isolation as long as the cross section is smaller than $\sigma/m \simlt 10\, \mathrm{cm}^2/\mathrm{g}$. The tidal stripping effects play a central role in the change in the density profile of the SIDM subhalo across its orbit (Subsection~\ref{subsec:tidal_ev_SIDM}). 
When the initial subhalo density is set to be consistent with the $\Lambda$CDM prediction at $z\sim2$, the SIDM subhaloes do not undergo the gravothermal collapse over 10 Gyr in our simulations.

\vspace{6pt}

\item The orbit of SIDM subhaloes can be precisely predicted by a simple framework based on point-mass approximation incorporated with the dynamical friction \citep{1943ApJ....97..255C} and the ram-pressure-induced deceleration \citep{2018MNRAS.474..388K} (Subsection~\ref{subsec:eq_motion}).

\vspace{6pt}

\item The time evolution of SIDM subhalo masses can be also explained 
by a common method accounting for the mass loss due to tidal stripping and ram-pressure effects 
(Subsection~\ref{subsec:mass_loss}). Our N-body simulations need an effective mass loss rate of the tidal stripping (Eq.~\ref{eq:mass_loss_TS}) to depend on the self-interacting cross section $\sigma/m$, 
that is a new systematic effect in the prediction of SIDM subhaloes.

\vspace{6pt}

\item Our semi-analytic model can provide a reasonable fit to the simulation results for various cross sections (including a velocity-dependent scenario as in Eq.~\ref{eq:vdep_sigma_over_m}), subhalo orbits,
and initial subhalo density profiles. A typical uncertainty in the model prediction is 0.1-0.2 dex for the SIDM subhalo density profiles over 10 Gyr in a range of $\sigma/m \simlt 10\, \mathrm{cm}^2/\mathrm{g}$.

\end{enumerate}

Our semi-analytic model provides a simple, efficient, and physically-intuitive prediction 
of SIDM subhaloes, but it has to be revised in various aspects for applications to real data sets.
The model should include more realistic effects, such as 
baryonic effects in a MW-sized host halo, 
the gravothermal instability induced by tidal stripping effects,
cosmological mass accretion around the host halo, and gravitational interaction among subhaloes 
in the host (see Section~\ref{sec:limitations} for details).
We expect the model to be improved on a step-by-step basis 
with a use of cosmological N-body simulations as well as isolated N-body simulations including 
baryonic components in the host gravitational potential.
This is along the line of our ongoing study.

\section*{acknowledgements} 
The authors thank the anonymous referee for reading the paper carefully and providing thoughtful comments, many of which have resulted in changes to the revised version of the manuscript.
The authors also thank Kohei~Hayashi and Ayuki~Kamada for useful discussions about modelling of SIDM haloes at early stages of this work.
The authors are indebted to Camila Correa for giving us comments on our SIDM implementation.
This work is supported by MEXT/JSPS KAKENHI Grant Numbers (19K14767, 19H01931, 20H05850, 20H05861,
21H04496).
Numerical computations were in part carried out on Cray XC50 at Center 
for Computational Astrophysics, National Astronomical Observatory of Japan, 
Oakforest-PACS at the CCS, University of Tsukuba, 
and  the computer resource offered under the category of General Project by 
Research Institute for Information Technology, Kyushu University.

\section*{Data Availability}
The data underlying this article will be shared on reasonable request to the corresponding author.



\bibliographystyle{mnras}
\bibliography{refs} 



\appendix

\section{Convergence tests for N-body simulations}\label{apdx:convergence_tests}

We here summarise convergence tests of our N-body simulations for halo-subhalo mergers.
In this Appendix, we work on the same parameter sets as ``SIDM1'' in Table~\ref{tab:sim_params}.
We run three different N-body simulations with the particle mass of 
$m_\mathrm{part}$ being 
$10^{4}$, $10^{5}$ and $10^{6}\, M_\odot$, respectively.
In each simulation, we set the gravitational softening length as in Eq~(\ref{eq:soften_length}).
Note that the host halo (subhalo at $t=0$) can be resolved with 
$10^{8} (10^{5})$, $10^{7} (10^{4})$, and $10^{6} (10^{3})$
when we set $m_\mathrm{part} = 10^{4}$, $10^{5}$ and $10^{6} M_\odot$.

\begin{figure}
 \includegraphics[width=\columnwidth]{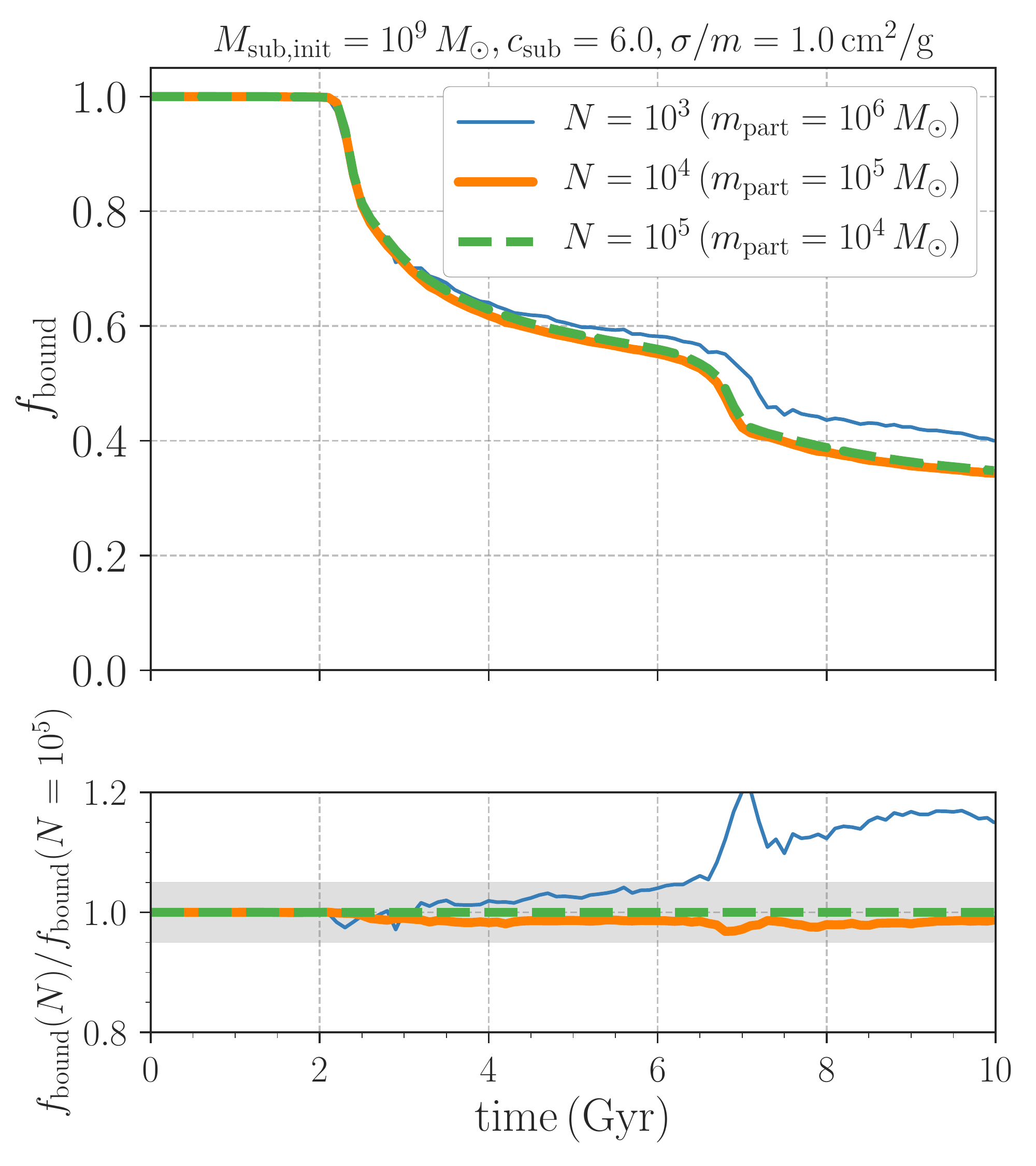}
 \caption{Convergence tests for evolution of subhalo bound mass.
 The top panel shows the fraction of subhalo bound mass (normalised to unity at $t=0$)
 when we vary the particle resolution in our simulations. 
 The bottom panel represents the fractional difference among the simulation results.
 The grey shaded region in the bottom shows a $\pm5\%$ difference.
 In each panel, the blue thin, orange thick, green dashed lines stand for the simulation results with $m_\mathrm{part} = 10^{6}$, $10^{5}$ and $10^{4}\, M_\odot$, respectively.
 This figure highlights that our fiducial run with $m_\mathrm{part} = 10^{5}\, M_\odot$ shows a converged result within a few percents. 
 }
 \label{fig:convergence_test1}
\end{figure}

\begin{figure}
 \includegraphics[width=\columnwidth]{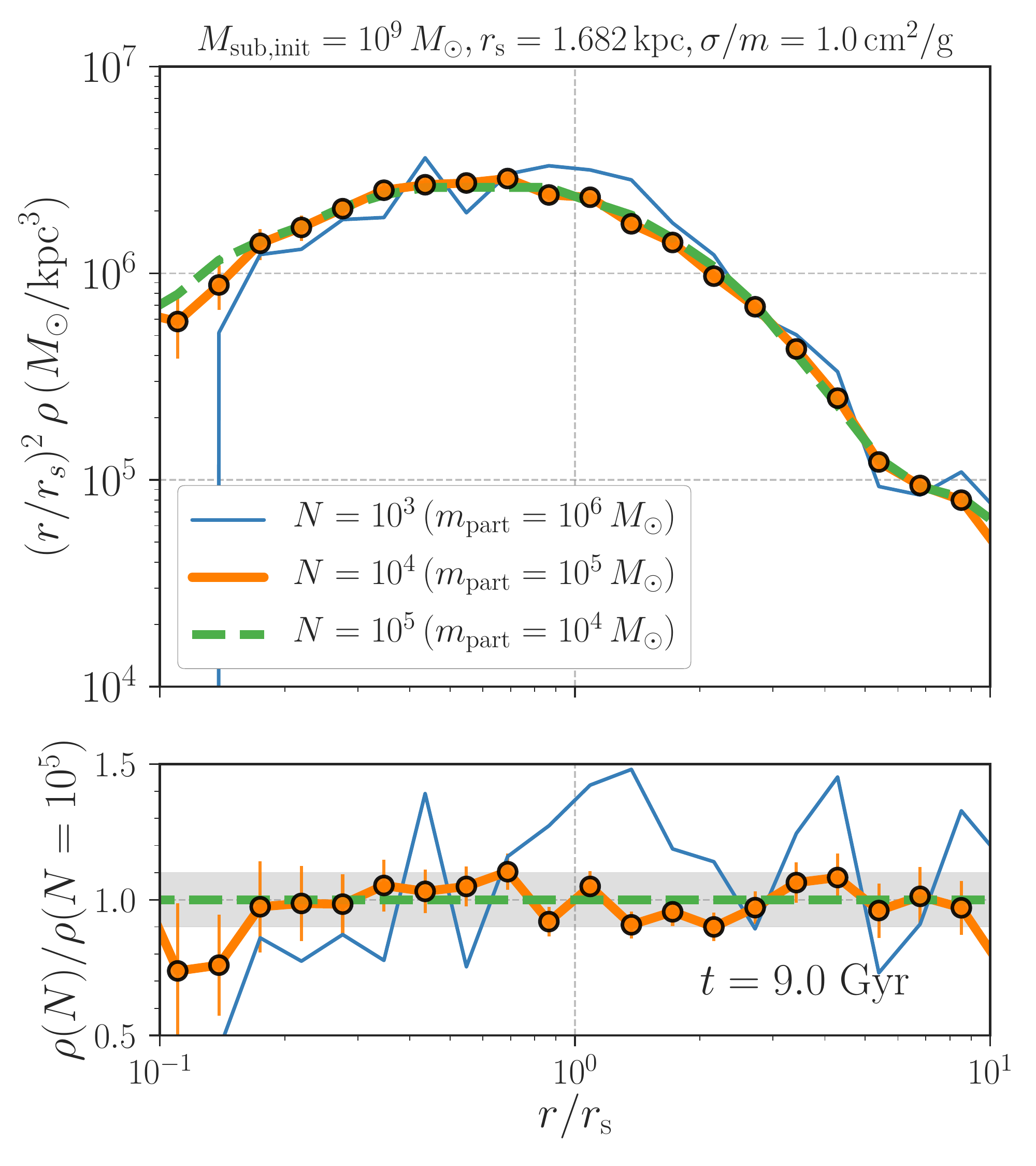}
 \caption{Convergence tests for subhalo density profiles.
 Similar legends are applied as in Figure~\ref{fig:convergence_test1}.
 The top panel shows the subhalo density profiles evolved by 9 Gyr when the resolution 
 is varied, while the bottom represents the fractional difference. 
 The grey region in the bottom panel highlights a $\pm10\%$ difference.
 }
 \label{fig:convergence_test2}
\end{figure}

Figures~\ref{fig:convergence_test1} and \ref{fig:convergence_test2}
summarise the convergence tests in our N-body simulations.
We found that our fiducial set up with $m_\mathrm{part}=10^{5}\, M_\odot$
can make the subhalo mass evolution converged within a $2-3\%$ level, 
while the subhalo density profile at $r/r_s \simgt 0.2$ in our simulations 
looks converged with a $10\%$-level precision.
We caution that the inner subhalo profile 
may suffer from some numerical resolution effects in our simulation sets.

\section{A test of SIDM implementation}\label{apdx:test_robertson}

\begin{figure}
 \includegraphics[width=\columnwidth]{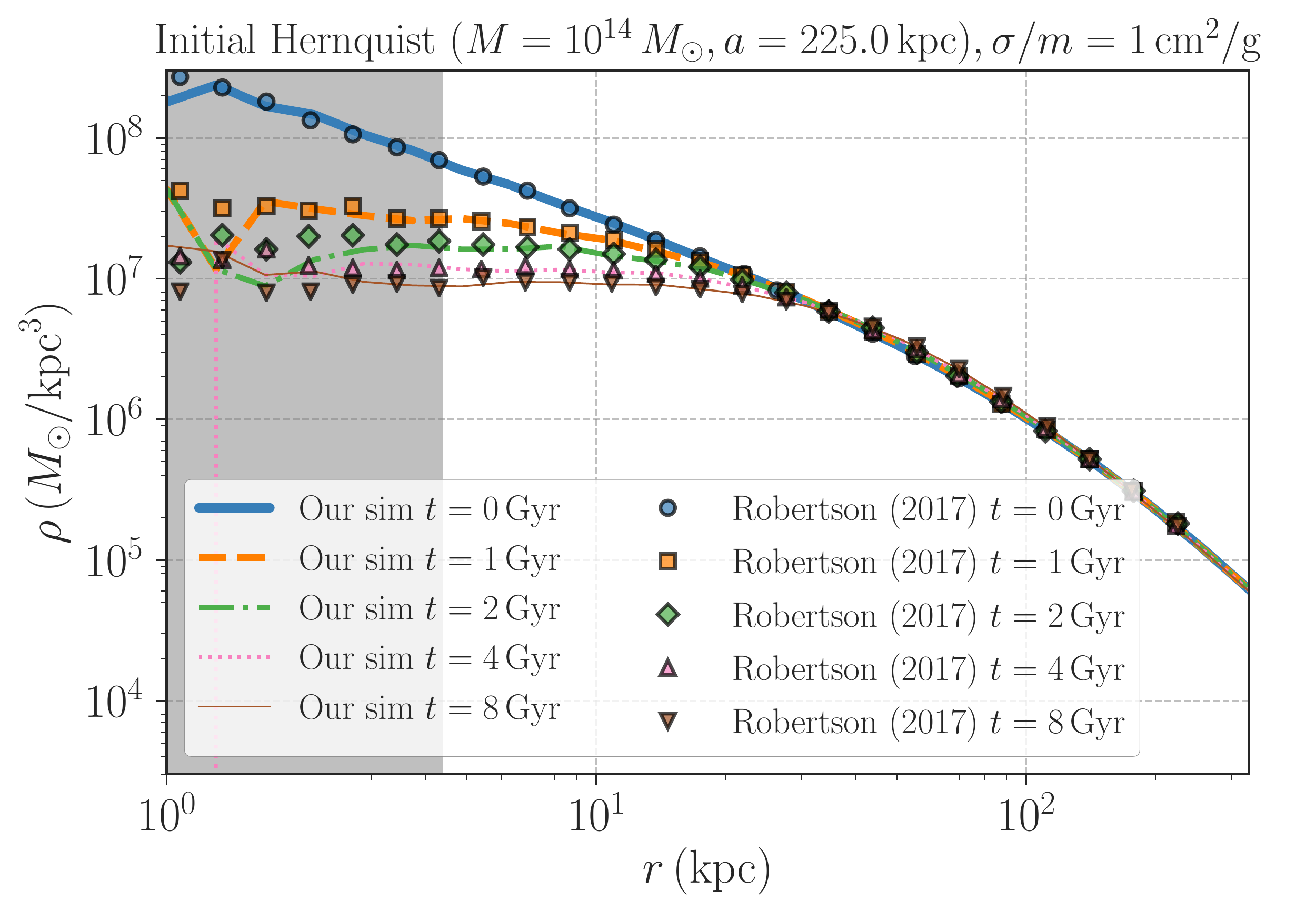}
 \caption{The halo core formation and size evolution for a Hernquist halo with $M=10^{14}\, M_\odot$ and the scaled radius $a=225\, \mathrm{kpc}$. In this figure, we consider the SIDM model with $\sigma/m=1\, \mathrm{cm}^2/\mathrm{g}$. The points show the result reported in \citet{2017PhDT.......206R}. The different lines represent our simulation results, demonstrating that our SIDM implementation provides a consistent time evolution of SIDM cores. The gray region highlights scales shorter than the gravitational softening length of $4.4\, \mathrm{kpc}$.
 }
 \label{fig:check_Hernquist_SIDM}
\end{figure}

\rev{As a test, we consider an isolated halo following a Hernquist profile at its initial state. The Hernquist profile is expressed as
\beqa
\rho(r) = \frac{M}{2\pi} \frac{a}{r(r+a)^3},
\eeqa
where $a$ is the scaled radius. For the initial Hernquist profile, we adopt the same parameters as in \citet{2017PhDT.......206R}. 
To be specific, we set the mass parameter of $10^{14}\, M_\odot$ and the scaled radius of $225\, \mathrm{kpc}$.
We ran the simulation with $128^3$ N-body particles, the gravitational softening length of $4.4\, \mathrm{kpc}$, and the cross section of $\sigma/m=1\, \mathrm{cm}^2/\mathrm{g}$. Note that those simulation parameters are also same as in \citet{2017PhDT.......206R}. For comparison, we extract the data points of SIDM density profiles from Figure 4.9 in \citet{2017PhDT.......206R} by using this website\footnote{\url{https://apps.automeris.io/wpd/}}. Figure~\ref{fig:check_Hernquist_SIDM} summarises the comparison of the halo core evolution for the Hernquist halo in our simulation with the results in \citet{2017PhDT.......206R}. 
We confirm that our SIDM implementation provides a good fit to the results 
in the literature.}

\section{Calibration of gravothermal fluid model for an isolated halo}\label{apdx:gravthermal_NFW}

In this Appendix, we describe our calibration of the gravothermal fluid model.
For the calibration, we perform N-body simulations of an isolated halo with 
its initial density profile following a NFW profile 
as varying the self-interacting cross section $\sigma/m$.
In these isolated simulations, we set the halo mass and the scaled radius at $t=0$
to be $10^{12}\, M_\odot$ and $r_s = 21.18\, \mathrm{kpc}$.
We examine five cross sections of $\sigma/m = 0.3, 1, 3, 10$ and $30\, \mathrm{cm}^2/\mathrm{g}$ and evolve the halo by 10 Gyr in our simulations.
The simulation outputs are stored with a time interval of $0.1\, \mathrm{Gyr}$,
producing 100 snapshots for a given SIDM model.
We refer the readers to Subsection~\ref{subsec:inital_condition} about how to prepare 
an isolated NFW halo.

\begin{figure}
 \includegraphics[width=\columnwidth]{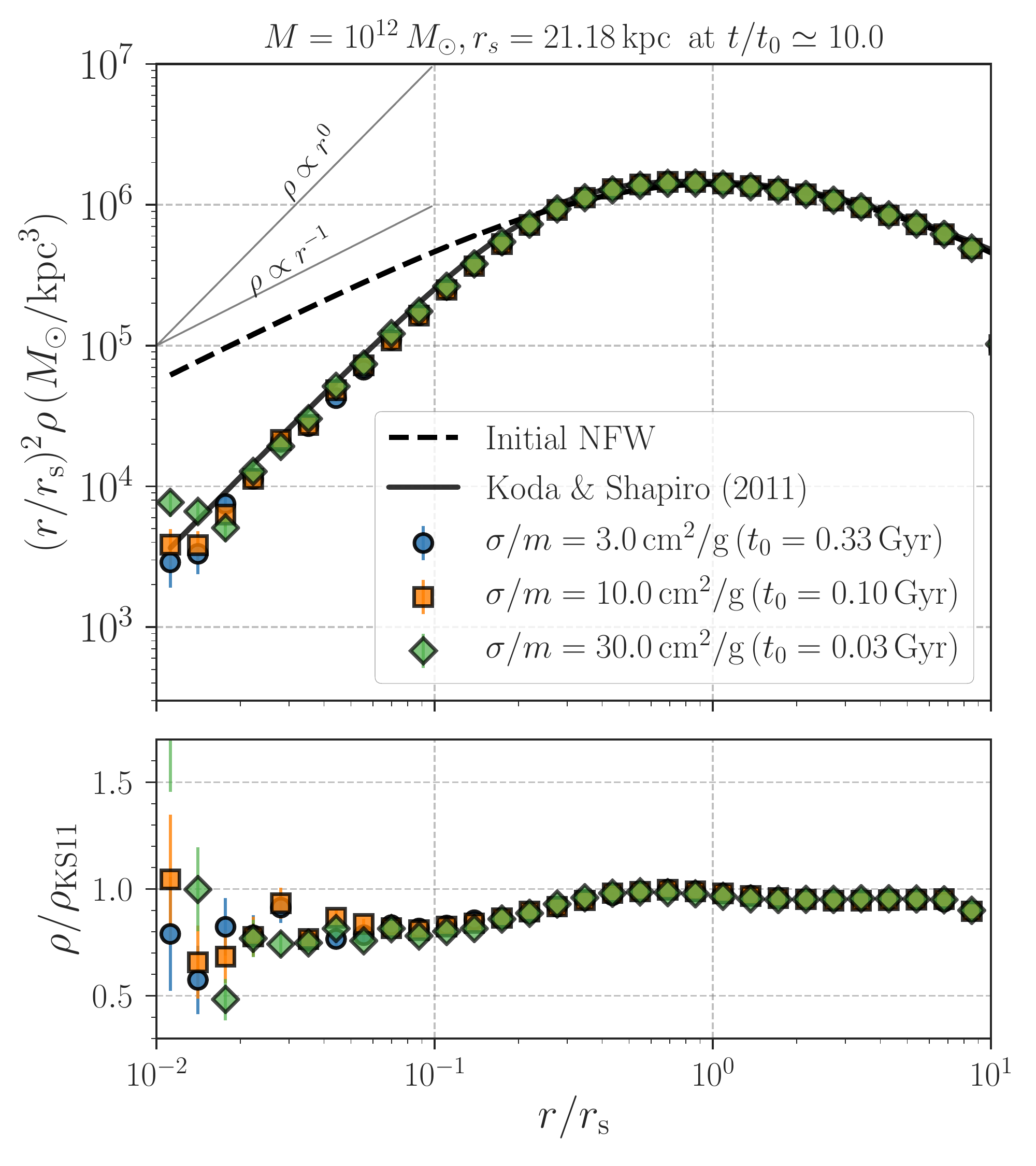}
 \caption{SIDM density profiles as a function of the cross section $\sigma/m$.
 In the upper panel, the solid line shows the gravothermal fluid model with the heat conductivity calibrated in \citet{2011MNRAS.415.1125K}, while different coloured symbols represent our N-body simulation results for an isolated halo with its mass of $10^{12}\, M_\odot$. We show the simulation results at a dimensionless time $t/t_0 =10$,
 where $t_0$ is a characteristic relaxation time scale given in Eq.~(\ref{eq:t_0_SIDM}).
 Because the time scale $t_0$ depends on $\sigma/m$, each symbol represents the density profile at different epoch; The blue circle shows the density profile at $t\simeq 3.3\, \mathrm{Gyr}$ for the SIDM with $\sigma/m = 3\, \mathrm{cm}^2/\mathrm{g}$,
 while the orange square and green diamond show the counterparts at 
 $t\simeq 1.0\, \mathrm{Gyr}$ and $t\simeq 0.3\, \mathrm{Gyr}$
 for $\sigma/m = 10\, \mathrm{cm}^2/\mathrm{g}$ and $30\, \mathrm{cm}^2/\mathrm{g}$, respectively.
 Note that the upper panel shows the quantity of $\sim r^2 \rho$.
 For ease of comparisons, we also show the initial NFW profile by the dashed line in the top panel. The bottom panel shows the fractional difference between the gravothermal fluid model and the simulation results at $t/t_0 = 3$, highlighting that a universal correction can be applied to the gravothermal fluid prediction for various $\sigma/m$.
 }
 \label{fig:host_density_normt}
\end{figure}

Figure~\ref{fig:host_density_normt} summarises the comparison of the SIDM density profile
between the simulation results and the gravothermal 
fluid model in \citet{2011MNRAS.415.1125K}.
In the figure, we show the density profiles at a dimensionless epoch $t/t_0=10$,
where $t_0$ is given by Eq.~(\ref{eq:t_0_SIDM}).
Once considering evolution with respect to dimensionless epochs $t/t_0$, 
we find that the gravothermal fluid model predicts almost an identical density profile at a given $t/t_0$ regardless of the exact value of $\sigma/m$.
The gravothermal fluid prediction is shown by the solid line in the top panel of figure~\ref{fig:host_density_normt}, while different coloured symbols represent
our simulation results at $t/t_0=10$.
Although the simulation results exhibit a $O(10)\%$ difference from the gravothermal fluid model at $r/r_s\simeq 0.1$, the difference is found to be almost independent on $\sigma/m$
if comparing the density profiles at the same dimensionless epoch $t/t_0$.
This finding motivates us to develop a correction function of the gravothermal fluid model below;
\beqa
\rho_\mathrm{SIDM}(r, t, \sigma/m) = \rho_\mathrm{gt}(r, t, \sigma/m) \, {\cal C}(r/r_s, t/t_0),
\eeqa
where ${\cal C}$ represents the correction function which we would like to find.
After some trials, we find that our simulation results can be well explained 
by a two-parameter function below;
\beqa
{\cal C}(x, \tilde{t}) = \frac{x^{\beta}+(1/2)^{\beta}}{(x+\gamma)^{\beta}},
\eeqa
where $x=r/(0.1 r_s)$ and we assume that $\beta$ and $\gamma$ depend on $\tilde{t}=t/t_0$.

\begin{figure*}
 \includegraphics[width=2\columnwidth]{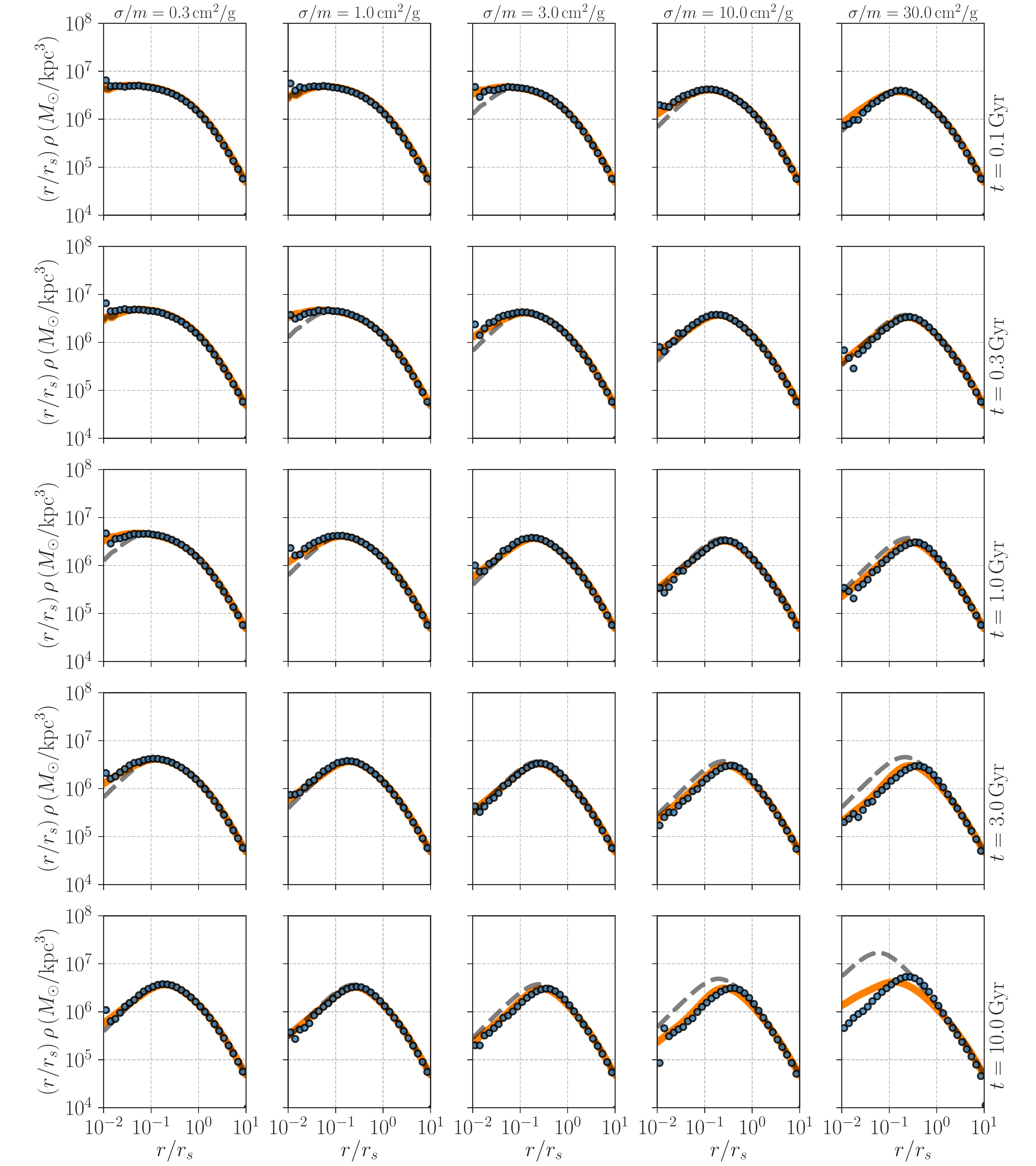}
 \caption{Tests of our calibrated gravothermal fluid model (Eq.~\ref{eq:final_iso_density}) against the N-body simulations of an isolated halo.
 In each panel, the blue circles show the simulation results, while
 the grey dashed and orange solid lines represent the model in \citet{2011MNRAS.415.1125K} and our calibrated model, respectively. 
 From left to right (top to bottom), 
 we show the comparisons as increasing $\sigma/m$ (epoch $t$).
 }
 \label{fig:test_gv_model}
\end{figure*}

Using the density profile of the simulated halo at a given snapshot and cross section of $\sigma/m$, we find the best-fit parameters of $\beta$ and $\gamma$
by minimising the chi-square value of
\beqa
\chi^2 = \sum_i \left[\rho_\mathrm{sim}(r_i, t, \sigma/m) - \rho_\mathrm{SIDM}(r_i, t, \sigma/m)\right]^2,
\eeqa
where $\rho_\mathrm{sim}$ represents the density profile of the simulated halo
and $r_i$ is the $i$-th bin in the halo-centric radius.
For this chi-square analysis, we perform a logarithmic binning in $r/r_s$ with 
the number of bins being 35 in a range of $0.01 < r/r_s < 30$ 
when computing the spherical density profile of the simulated halo.
After finding the best-fit parameters for a given set of 
snapshot time $t$ and cross section $\sigma/m$, we derive the $t/t_0$-dependence as in Eqs.~(\ref{eq:gv_param3}) and (\ref{eq:gv_param4}).
Figure~\ref{fig:test_gv_model} summarises our calibration, demonstrating
that the model of Eq.~(\ref{eq:final_iso_density}) can provide a good fit to
the simulation results for a wide range of $\sigma/m$ and $t$.
We confirm that our calibrated model has a $10\%$-level precision 
in the range of $t/t_0 \simlt 100$.
It would be worth noting that our model has been calibrated for a specific initial condition.
Hence, our model can not be applied to general cases, but it would provide a reasonable fit to the SIDM density profile as long as its initial density follows a NFW profile.
\rev{A caveat is that our calibration may depend on a choice of boundary radius in an isolated SIDM halo as discussed in \citet{2011MNRAS.415.1125K}. Note that the model in \citet{2011MNRAS.415.1125K} has been calibrated with simulation results assuming the halo boundary radius is set to 100 times as large as the NFW scaled radius, while we adopted a more realistic situation (i.e. the halo concentration of 10). We leave it to investigate possible effects of the halo boundary radii in SIDM simulations for future studies.}

\section{A fitting formula of the transfer function for tidally truncated density profiles}\label{apdx:tranfer_func}

In this appendix, we provide a fitting formula of the transfer function developed in \citet{2019MNRAS.490.2091G}. In the context of tidal evolution of collision-less dark matter subhaloes, the transfer function is commonly defined as
\beqa
H(r, t) = \frac{\rho(r,t)}{\rho(r,t=0)},
\eeqa
where $H$ is the transfer function, $r$ is the radius from the centre of the subhalo, 
and $\rho(r,t)$ is the subhalo density profile at an epoch of $t$.
Using a set of collision-less N-body simulations of minor mergers, 
\citet{2019MNRAS.490.2091G} found that $H$ can be well approximated as the form below;
\beqa
H(r, t) = \frac{f_\mathrm{te}}{1+
\left(\tilde{r}\left[\frac{\tilde{r}_\mathrm{sub,vir}-\tilde{r}_\mathrm{te}}{\tilde{r}_\mathrm{sub,vir}\tilde{r}_\mathrm{te}}\right]\right)^{\delta}}, \label{eq:transfer_func_GB19}
\eeqa
where $\tilde{r} = r/r_\mathrm{s, init}$ such that all radii in Eq.~(\ref{eq:transfer_func_GB19}) are normalized to the initial NFW scale radius of the subhalo $r_\mathrm{s,init}$.

Eq.~(\ref{eq:transfer_func_GB19}) contains three model parameters and those depend on
the initial subhalo concentration $c_\mathrm{sub}$ and the bound mass fraction of the subhalo at the epoch $t$ (denoted as $f_\mathrm{bound}$).
Throughout this paper, we adopt 
\beqa
f_\mathrm{te} &=& f^{a_1 (c_\mathrm{sub,10})^{a_2}}_\mathrm{b} \, c^{a_3(1-f_\mathrm{bound})^{a_4}}_\mathrm{sub}, \\
\tilde{r}_\mathrm{te} &=& \tilde{r}_\mathrm{sub,vir}\, 
f^{b_1 (c_\mathrm{sub,10})^{b_2}}_\mathrm{b} \, c^{b_3(1-f_\mathrm{bound})^{b_4}}_\mathrm{sub}\, \nonumber \\
&&
\quad \quad \quad \quad \quad \quad
\times
\exp\left[b_5 (c_\mathrm{sub,10})^{b_6}(1-f_\mathrm{bound})\right], \\
\delta &=& c_0\, f^{c_1 (c_\mathrm{sub,10})^{c_2}}_\mathrm{b} \, c^{c_3(1-f_\mathrm{bound})^{c_4}}_\mathrm{sub},
\eeqa
where $c_\mathrm{sub,10}=c_\mathrm{sub}/10$,
$a_1=0.338$, $a_2=0.000$, $a_3=0.157$, $a_4=1.337$,
$b_1=0.448$, $b_2=0.272$, $b_3=-0.199$, $b_4=0.011$, $b_5=-1.119$, $b_6=0.093$,
$c_0=2.779$, $c_1=-0.035$, $c_2=-0.337$, $c_3=-0.099$, and $c_4=0.415$.
Note that the function in Eq.~(\ref{eq:transfer_func_GB19}) has been calibrated for the collision-less dark matter. Hence, we have tested if it can be applied to collisional scenarios
in Subsection~\ref{subsec:tidal_ev_SIDM}.

\section{A semi-analytic model in Jiang~et~al. (2021a)}\label{apdx:model_J21}

For the sake of clarity, we here summarise a semi-analytic model in \citet{2021arXiv210803243J}.
The model assumes that an isolated SIDM halo follows a NFW profile at its initial state
and the density profile at a given age $t$ can be approximated as
\beqa
\rho_\mathrm{SIDM, J21}(r) &=& \frac{1}{4\pi r^2}\frac{\mathrm{d}M_\mathrm{SIDM,J21}}{\mathrm{d}r}, \\
M_\mathrm{SIDM,J21}(r) &=& \tanh\left(\frac{r}{r_c}\right) \, M_\mathrm{NFW}(r),
\eeqa
where $M_\mathrm{NFW}(r)$ is the enclosed mass of the initial NFW profile,
and $r_c$ represents an effective core radius of the SIDM halo
and depends on the time of $t$.
To be specific, $r_c$ is given by $\mathrm{min}[0.5r_1, r_s]$ ($r_s$ is the scaled radius for the initial NFW profile) and $r_1$ is set by
\beqa
\langle \sigma v /m\rangle \, \rho_\mathrm{SIDM,J21}(r_1) \, t = 1, \label{eq:r_1}
\eeqa
where the above equation means that the SIDM core size can be related to the radius 
where every SIDM particle has interacted once by the time $t$.
The average in Eq.~(\ref{eq:r_1}) is given by 
\beqa
\langle \sigma v/m  \rangle = \int_0^{\infty}\mathrm{d}v\, v\, \frac{\sigma}{m}
\, f(v;v_c),
\eeqa
where $f(v;v_c)$ is the Maxwell-Boltzmann distribution of Eq.~(\ref{eq:MB_dist}).
The parameter $v_c$ is set to $4\, \sigma_{v, \mathrm{J21}}(r)/\sqrt{\pi}$ with
\beqa
\sigma^2_{v, \mathrm{J21}}(r) = \frac{1}{\rho_\mathrm{SIDM, J21}(r)}
\int_r^{\infty} \mathrm{d}r'\, \frac{\rho_\mathrm{SIDM,J21}(r')}{r'}
\frac{G M_\mathrm{SIDM, J21}(r')}{r'}.
\eeqa

In \citet{2021arXiv210803243J}, the authors solve 
the orbital evolution of infalling subhaloes 
as same as in Subsection~\ref{subsec:eq_motion}.
The mass loss due to the tidal stripping is also set by Eq.~(\ref{eq:mass_loss_TS}),
but they adopt ${\cal A}=0.55$ and $q=1$ for any SIDM models.
They also take into account the mass loss by the self-interacting evaporation as in Eq.~(\ref{eq:mass_loss_RPe}).
For a given mass loss rate, the model in \citet{2021arXiv210803243J}
then updates the subhalo density profile after a finite time of $\Delta t$ 
by rules below;
\beqa
M_\mathrm{sub}(\rho_0, r'_\mathrm{out}) - M_\mathrm{sub}(\rho_0, r_\mathrm{out})
&=& \left(\frac{\mathrm{d}M_\mathrm{sub}}{\mathrm{d}t}\right)_\mathrm{TS}\Delta t, \label{eq:mass_sub1}\\
M_\mathrm{sub}(\rho'_0, r_\mathrm{out}) - M_\mathrm{sub}(\rho_0, r_\mathrm{out})
&=& \left(\frac{\mathrm{d}M_\mathrm{sub}}{\mathrm{d}t}\right)_\mathrm{RPe}\Delta t,
\label{eq:mass_sub2}
\eeqa
where $M_\mathrm{sub}(\rho_0, r_\mathrm{out})$ is the enclosed mass of the subhalo at its boundary radius of $r_\mathrm{out}$ with the density amplitude being $\rho_0$.
We denote $r'_\mathrm{out}$ and $\rho'_0$ as the quantities to be updated.
Eqs.~(\ref{eq:mass_sub1}) and (\ref{eq:mass_sub2}) are designed so that
the tidal stripping can remove the subhalo mass at its outermost radius,
while the ram-pressure effects can affect the overall subhalo density profile.


\bsp	
\label{lastpage}
\end{document}